\documentclass[12pt,a4paper]{article}
\usepackage{graphicx} 
\usepackage{amsmath}
\usepackage{amssymb}
\usepackage[shortlabels]{enumitem}
\setlist{noitemsep}

\usepackage[left=2.0cm, right=2.0cm]{geometry}
\usepackage[square,numbers]{natbib}

\title{Quantum three-body problem for nuclear physics}
\author{Emile Meoto \vspace{0.5cm} \\  Department of Physics, University of Buea \\ P. O. Box 63 Buea, South West Region, Cameroon \vspace{0.5cm} \\ Email: meoto.emile@ubuea.cm or emeotoson@gmail.com}
\date{\today}

\begin{document}

\maketitle

\begin{abstract}
A brief excursion into the three-body problem in quantum mechanics is presented for graduate students or researchers in nuclear physics. Starting from single-particle coordinates, the three-body Schr\"{o}dinger equation is systematically transformed into a representation in Jacobi coordinates. Gradient, Laplacian, and kinetic energy operators are explicitly derived using the multivariable chain rule. Faddeev equations are reformulated in hyperspherical coordinates. In all transformations (from single-particle coordinates to Jacobi coordinates, rotation between Jacobi coordinates and from Jacobi coordinates to hyperspherical coordinates) the determinant of the Jacobian matrix is computed to ensure correct transformation of volume elements. The Faddeev equations in hyperspherical coordinates are projected onto a hyperspherical harmonics basis, leading to the coupled hyperradial equations that define the hyperspherical harmonics method.  
\end{abstract}

\section{Introduction}

In quantum mechanics, the three-body problem refers to a system of three bodies interacting with each other through forces that depend on the specific nature of these particles. The dynamics of the system is governed by the Schr\"{o}dinger equation or any equivalent formalism. Examples of interaction forces include the nuclear force, the Coulomb force and van der Waals forces. This should be contrasted with the classical three-body problem that deals with three bodies that interact with each other through the gravitational force, and their dynamics is described by Newtonian mechanics or analytical mechanics (Lagrangian or Hamiltonian formulation). The quantum-mechanical three-body problem arises in many disciplines. Examples in nuclear physics, atomic and molecular physics, and particle physics are presented here. For each example, the three interacting bodies are identified. 

\begin{enumerate}[(1)]
\item Nuclear physics
              \begin{enumerate}[(i)]
               \item Triton (nucleus of tritium i.e. hydrogen-3): one proton and two neutrons.
               \item Hypertriton: a lambda baryon ($\Lambda$), a proton and a neutron.
               \item Helium-3 nucleus: two protons and one neutron.
               \item Kaonic systems: one kaon and two protons.
              \end{enumerate}
    \item Atomic and molecular physics
           \begin{enumerate}[(i)]
               \item Helium atom: a helium nucleus with its two electrons.  
               \item Muonic helium atom: a helium nucleus, an electron and a muon. 
               \item Negative hydrogen ion or a hydride ion ($H^{-}$): two electrons and one proton.
               \item Hydrogen molecular ion ($H^{+}_2$): two hydrogen nuclei and one electron.
               \item Rydberg atoms with core electrons: nucleus, Rydberg electron and the core electrons.
               \item Three-body recombination in ultracold atoms: three atoms.
               \item Efimov states: three atoms. 
               \item Muon-catalyzed fusion: two deuterium (or tritium) nuclei and one muon. 
           \end{enumerate}
      \item Particle physics
              \begin{enumerate}[(i)]
               \item Proton: two up quarks and a down quark. 
               \item Neutron: one up quark and two down quarks.
               \item Lambda baryon: one up quark, one down quark and one strange quark.
           \end{enumerate}  
\end{enumerate}

The three-body problem is a special case of the few-body problem. For a few-body problem, the number of interacting bodies is usually small, typically six or fewer. Other examples of few-body problems are the four-body problem, the five-body problem and the six-body problem. In such systems, the dynamics of each particle is explicitly determined. When the number of interacting bodies becomes large, it is called a many-body problem. The boundary between few-body and many-body physics can sometimes be fluid. In the many-body problem, the explicit treatment of the dynamics of individual particles is no longer feasible or of primary interest. Instead, the focus shifts to understanding how interactions between particles give rise to collective or emergent behaviour. The many-body problem is investigated through statistical methods such as mean-field theories, Monte Carlo techniques, Density Functional Theory, and related approaches.

Two aspects are of central importance in any few-body problem: whether it is integrable and whether it is exactly solvable. In quantum mechanics, a system with $N$ degrees of freedom is said to be integrable if it possesses $N$ independent conserved quantities (constants of motion) whose corresponding operators commute mutually and with the Hamiltonian. These operators correspond to conserved quantities whose eigenvalues uniquely label the eigenstates. For example, the two-body problem is integrable, whereas the three-body problem is, in most cases, non-integrable. A problem is exactly solvable if its solution can be expressed in closed form using analytic functions. A system may be integrable yet still not exactly solvable. For instance, the two-body problem is integrable: it is exactly solvable for a 3D harmonic oscillator potential but not for a Woods–Saxon potential.  

This paper presents a comprehensive and fully explicit derivation of the quantum-mechanical three-body Schr\"odinger and Faddeev formalisms, in both Jacobi and hyperspherical coordinate systems. Particular emphasis is placed on the rigorous transformation of differential operators and the careful evaluation of Jacobian determinants governing coordinate changes and volume elements.
Starting from single-particle coordinates, all transformations are carried out step by step using the multivariable chain rule, leading to transparent expressions for gradient and Laplacian operators and to a clear demonstration of the separation between internal and centre-of-mass dynamics. In contrast to many existing reviews and pedagogical treatments, where some key mathematical steps are often omitted or presented only schematically, this work provides complete derivations of the Jacobian determinants, operator representations in various coordinate systems, and hyperspherical projections required for the hyperspherical harmonics expansion and the construction of coupled hyperradial equations. 

\section{Three-body time-independent Schr\"{o}dinger equation}

Consider a three-body problem in which the bodies have masses $m_1$, $m_2$ and $m_3$. Let the bodies have Cartesian position vectors $\vec{r}_1$, $\vec{r}_2$ and $\vec{r}_3$, respectively, with respect to a reference frame fixed to the laboratory. The dynamics of this system is described by the three-body time-independent Schr\"{o}dinger equation, which is given by

\begin{eqnarray}
H \phi (\vec{r}_1, \vec{r}_2, \vec{r}_3) = E \phi(\vec{r}_1, \vec{r}_2, \vec{r}_3), 
\end{eqnarray}

where $H$ is the Hamiltonian of the system. Written explicitly, the equation reads

\begin{eqnarray}
(T + V(\vec{r}_1, \vec{r}_2, \vec{r}_3)) \phi (\vec{r}_1, \vec{r}_2, \vec{r}_3) = E \phi(\vec{r}_1, \vec{r}_2, \vec{r}_3).  
\end{eqnarray}

Here $T$ is the kinetic energy operator and $V(\vec{r}_1, \vec{r}_2, \vec{r}_3)$ is the potential that describes the interaction between the bodies. In most cases of relevance to nuclear physics, this potential is the sum of pairwise interactions $V_{ij}$. In that case the three-body Schr\"{o}dinger equation can further be written as

\begin{eqnarray}
\label{eq:schrodinger}
\left \{ - \sum_{i=1}^{3} \frac{\hbar^2}{2m_i} \nabla^2_{\vec{r}_i}  + \sum_{i<j} V_{ij}(\vec{r}_i, \vec{r}_j) \right \} \phi(\vec{r}_1, \vec{r}_2, \vec{r}_3) = E \phi(\vec{r}_1, \vec{r}_2, \vec{r}_3)  
\end{eqnarray}

The condition $i<j$ in the potential sum guarantees that each pair interaction is counted only once, since $V_{ij}=V_{ji}$ (i.e. $V_{12}=V_{21}$, $V_{23}=V_{32}$ and $V_{13}=V_{31}$). Writing $\vec{r}_i=(x_i, y_i, z_i)$, the Laplacian for particle $i$ is given explicitly as 
\begin{eqnarray}
\nabla^2_{\vec{r}_i} = \frac{\partial^2}{\partial x^2_{i}} + \frac{\partial^2}{\partial y^2_{i}} + \frac{\partial^2}{\partial z^2_{i}}
\end{eqnarray}

Since the potential is the sum of pairwise interactions, single-particle coordinates ($\vec{r}_1$, $\vec{r}_2$ and $\vec{r}_3$) are not well suited for describing the correlations of the system. This is because the interactions depend on the relative distances between particles rather than their absolute positions. Consequently, relative coordinates are a more natural choice.

\subsection{Jacobi coordinates: Relative and centre-of-mass coordinates}

The three-body Schr\"{o}dinger equation in single-particle coordinates, as given by Eq. \ref{eq:schrodinger}, is 9-dimensional. We now introduce generalised relative and centre-of-mass coordinates called Jacobi coordinates, which are constructed from the single-particle coordinates ($\vec{r}_1$, $\vec{r}_2$ and $\vec{r}_3$). Jacobi coordinates ($\vec{\eta}_i$, $\vec{\lambda}_i$, $\vec{R}$) are then defined as follows:

\begin{subequations}\label{eq:jacobi_defined}
\begin{align} 
\vec{\eta}_i &= \sqrt{\frac{\mu_{jk}}{m}}  \left(\vec{r_j} -  \vec{r_k} \right) \\
\vec{\lambda}_i &= \sqrt{\frac{\mu_{i(jk)}}{m}}   \left( \vec{r_i} - \frac{m_j \vec{r_j} + m_k \vec{r_k}}{m_j + m_k} \right) \\
\vec{R} &= \frac{m_i \vec{r_i} + m_j \vec{r_j} + m_k \vec{r_k}}{m_i + m_j + m_k}
\end{align}
\end{subequations}

where $m$ is an arbitrary reference mass and 

\begin{subequations}
\begin{align}
\frac{1}{\mu_{jk} } &= \frac{1}{m_j} + \frac{1}{m_k}  \\
\frac{1}{\mu_{i(jk)} } &= \frac{1}{m_i} + \frac{1}{m_j + m_k}
\end{align}
\end{subequations}

Here, $\mu_{jk}$ is the reduced mass of the interacting pair $(j,k)$ while $\mu_{i(jk)}$ is the reduced mass of the spectator $i$ and the interacting pair $(j,k)$. These scaling factors may be written explicitly as

\begin{subequations}
\begin{align}
\mu_{jk} &= \frac{m_j m_k}{m_j + m_k} \\
\mu_{i(jk)} &=\frac{m_i (m_j + m_k)}{m_i + m_j+ m_k}
\end{align}
\end{subequations}

The indices $i$, $j$ and $k$ run in a cyclic manner through the triple (1,2,3). The last Jacobi coordinate is the centre-of-mass of the three bodies. The notation used here to write Jacobi coordinates is called spectator notation. This means one particle (the spectator) is singled out, while the remaining particles are treated as interacting with each other. The coordinate $\vec{\eta}_i $ refers to a configuration in which particle $i$ is the spectator while particles $j$ and $k$ are interacting. There are two common ways of writing the mass scales in Jacobi coordinates: hierarchical scaling and equal-mass scaling. The mass scales used here ($\mu_{jk}$ and $\mu_{i(jk)}$) are hierarchical scales. In equal-mass scaling, all Jacobi coordinates are scaled by the same factor, typically using the total mass of the system. 
The transformation from ($\vec{r}_1$, $\vec{r}_2$ and $\vec{r}_3$) to ($\vec{\eta}_i$, $\vec{\lambda}_i$ and $R$) is linear, and therefore can be written as follows:

\begin{align}
\begin{pmatrix}
\vec{\eta}_i \\
\vec{\lambda}_i \\
\vec{R}
\end{pmatrix}
= S
\begin{pmatrix}
\vec{r}_i \\
\vec{r}_j \\
\vec{r}_k
\end{pmatrix}.
\end{align}

The $9 \times 9$ matrix S is given explicitly as 
\begin{align}
S = 
\begin{pmatrix}
\mathbf{0}_{3\times 3} & +\sqrt{\dfrac{\mu_{jk}}{m}} \mathbb{I}_3 & -\sqrt{\dfrac{\mu_{jk}}{m}} \mathbb{I}_3 \\[10pt]
\sqrt{\dfrac{\mu_{i(jk)}}{m}} \mathbb{I}_3 & -\sqrt{\dfrac{\mu_{i(jk)}}{m}}\dfrac{m_j}{m_j+m_k}\,\mathbb{I}_3 &
-\sqrt{\dfrac{\mu_{i(jk)}}{m}}\dfrac{m_k}{m_j+m_k}\,\mathbb{I}_3 \\[10pt]
\dfrac{m_i}{M}\,\mathbb{I}_3 & \dfrac{m_j}{M}\,\mathbb{I}_3 & \dfrac{m_k}{M}\,\mathbb{I}_3
\end{pmatrix},
\end{align}

where \(\mathbb{I}_3\) denotes the \(3\times 3\) identity matrix and $\mathbf{0}_{3\times 3}$ denotes the \(3\times 3\) zero matrix (this \(3\times 3\) zero block arises because $\vec{\eta}_i$ does not depend on $\vec{r}_i$). The Jacobian matrix of the transformation is given by

\begin{align} \label{eq:jacobian2}
J_1=\frac{\partial(\vec{r}_i, \vec{r}_j,\vec{r}_k)}{\partial (\vec{\eta}_i, \vec{\lambda}_i,\vec{R})}
\end{align}

In order to compute the Jacobian determinant of the transformation, one can invert the relations defining the Jacobi coordinates, so that the partial derivatives in Eq. \ref{eq:jacobian2} are computed. However, an easier approach is to compute the Jacobian determinant of the inverse transformation, since

\begin{align}
\left|\frac{\partial(\vec{r}_i, \vec{r}_j,\vec{r}_k)}{\partial (\vec{\eta}_i, \vec{\lambda}_i,R)} \right|=1/\left|\frac{\partial(\vec{\eta}_i, \vec{\lambda}_i,R)}{\partial (\vec{r}_i, \vec{r}_j,\vec{r}_k)} \right|
\end{align}

Therefore,

\begin{align}
\det(J_2) = \left|\frac{\partial(\vec{\eta}_i, \vec{\lambda}_i,R)}{\partial (\vec{r}_1, \vec{r}_2,\vec{r}_3)} \right|
= \begin{vmatrix}
\frac{\partial \vec{\eta}_i}{\partial \vec{r}_i} & \frac{\partial \vec{\eta}_i}{\partial \vec{r}_j} & \frac{\partial \vec{\eta}_i}{\partial \vec{r}_k} \\
\frac{\partial \vec{\lambda}_i}{\partial \vec{r}_i} & \frac{\partial \vec{\lambda}_i}{\partial \vec{r}_j} & \frac{\partial \vec{\lambda}_i}{\partial \vec{r}_k} \\
\frac{\partial R}{\partial \vec{r}_i} & \frac{\partial R}{\partial \vec{r}_j} & \frac{\partial R}{\partial \vec{r}_k}
\end{vmatrix},
\end{align}

The derivatives in the first, second and third rows are as follows:

\begin{align}
\frac{\partial \vec{\eta}_i}{\partial \vec{r}_i} = \mathbf{0}_{3\times 3}, \quad
\frac{\partial \vec{\eta}_i}{\partial \vec{r}_j} = +\sqrt{\frac{\mu_{jk}}{m}} \mathbb{I}_3, \quad
\frac{\partial \vec{\eta}_i}{\partial \vec{r}_k} = -\sqrt{\frac{\mu_{jk}}{m}} \mathbb{I}_3.
\end{align}

\begin{align}
\frac{\partial \vec{\lambda}_i}{\partial \vec{r}_i} = \sqrt{\frac{\mu_{i(jk)}}{m}} \mathbb{I}_3, \quad
\frac{\partial \vec{\lambda}_i}{\partial \vec{r}_j} = -\sqrt{\frac{\mu_{i(jk)}}{m}} \left( \frac{m_j}{m_j + m_k} \right) \mathbb{I}_3, \quad
\frac{\partial \vec{\lambda}_i}{\partial \vec{r}_k} = -\sqrt{\frac{\mu_{i(jk)}}{m}} \left( \frac{m_k}{m_j + m_k} \right) \mathbb{I}_3
\end{align}

\begin{align}
\frac{\partial \vec{R}}{\partial \vec{r}_i} = \frac{m_i}{M} \mathbb{I}_3, \quad
\frac{\partial \vec{R}}{\partial \vec{r}_j} = \frac{m_j}{M} \mathbb{I}_3, \quad
\frac{\partial \vec{R}}{\partial \vec{r}_k} = \frac{m_k}{M} \mathbb{I}_3.
\end{align}
where $M=m_i + m_j + m_k$. Using these first derivatives, the determinant of the $9 \times 9$ Jacobian matrix is computed as follows (common factors in each row have been factorised):

\begin{align}
\det(J_2)&= \left( \sqrt{\frac{\mu_{jk}}{m}}.\sqrt{\frac{\mu_{i(jk)}}{m}}.\frac{1}{M}
\begin{vmatrix}
0   &   1                     & -1            \\ 
1   & -\frac{m_j}{m_j + m_k}  & -\frac{m_k}{m_j + m_k}\\ 
m_i & m_j                     & m_k  
\end{vmatrix} 
\right)^3
\end{align}

The power of 3 arises because of the block structure of the Jacobian matrix.

\begin{align}
\det(J_2)&= \left( \sqrt{\frac{\mu_{jk}}{m}}.\sqrt{\frac{\mu_{i(jk)}}{m}}.\frac{1}{M} (-M) \right)^3 \notag \\
&= \left(- \sqrt{\frac{\mu_{jk}}{m}}.\sqrt{\frac{\mu_{i(jk)}}{m}} \right)^3 \notag \\
&= \left(-\sqrt{\frac{m_i m_j m_k}{m^2 M}} \right)^3 \notag \\
| \det(J_2) | &= \left( \frac{m_i m_j m_k}{m^2 M} \right)^{3/2} 
\end{align}

Therefore, the Jacobian determinant associated with the linear transformation from single-particle coordinates \((\vec r_i, \vec r_j, \vec r_k)\) to Jacobi coordinates
\((\vec \eta_i, \vec \lambda_i, \vec R)\) is given by

\begin{align}
\left| \det(J_1) \right|
&= 1/ \left| \det(J_2) \right| \notag \\
&= \left( \frac{m_i m_j m_k}{m^2 M} \right)^{-3/2} ,
\end{align}

This determinant represents the metric factor associated with the coordinate transformation and quantifies how infinitesimal volume elements are rescaled when
passing from single-particle coordinates to Jacobi coordinates. Consequently, the
volume elements transform as

\begin{align}
d^3 r_i \, d^3 r_j \, d^3 r_k
= \left| \det(J_1) \right| \,
d^3 \eta_i \, d^3 \lambda_i \, d^3 R .
\end{align}

The inclusion of this Jacobian factor is essential for preserving the normalization
of the wavefunction and for the consistent evaluation of observables in Jacobi
coordinates. In particular, the probability normalization condition takes the form
\begin{align}
\int
\left| \psi(\vec r_i, \vec r_j, \vec r_k) \right|^2
\, d^3 r_i \, d^3 r_j \, d^3 r_k
=
\int
\left| \psi(\vec \eta_i, \vec \lambda_i, \vec R) \right|^2
\left| \det(J_1) \right|
\, d^3 \eta_i \, d^3 \lambda_i \, d^3 R .
\end{align}

\subsection{Kinetic energy operator in Jacobi coordinates}

The kinetic energy operator in single-particle coordinates is now transformed to a kinetic energy operator in Jacobi coordinates. This is achieved by applying the chain rule for multivariable functions. In single-particle coordinates, the kinetic energy operator for a configuration in which the spectator is $i$ is given by 

\begin{align}
\label{eq:kinetic_single}
T=- \frac{\hbar^2}{2m_i} \nabla^2_{\vec{r}_i} - \frac{\hbar^2}{2m_j} \nabla^2_{\vec{r}_j} - \frac{\hbar^2}{2m_k} \nabla^2_{\vec{r}_k} 
\end{align}

The del operators $\nabla_{\vec{r}_i}$, $\nabla_{\vec{r}_j}$ and $\nabla_{\vec{r}_k}$ are now written as combinations of del operators in Jacobi coordinates ($\nabla_{\vec{\eta}_i}$, $\nabla_{\vec{\lambda}_i}$ and $\nabla_{\vec{R}}$). In this representation, the kinetic energy operator is written as a sum of off-diagonal and diagonal terms. This approach is motivated by the fact that the transformation to mass-scaled Jacobi coordinates, in most cases, is intended to diagonalise the kinetic energy operator. Separating the operator this way therefore makes it easier to verify this diagonalisation. Diagonalisation of the kinetic energy operator by transformation to Jacobi coordinates has been extensively discussed in \cite{for2010}.

\begin{enumerate}[(1)]
    \item For the $i$th particle ($\nabla_{\vec{r}_i}$) \\
    \begin{align}
     \nabla_{\vec{r}_i} = \frac{\partial \vec{\eta}_i }{\partial \vec{r}_i } \nabla_{\vec{\eta}_i} + \frac{\partial \vec{\lambda}_i }{\partial \vec{r}_i } \nabla_{\vec{\lambda}_i} + \frac{\partial \vec{R}_i }{\partial \vec{r}_i } \nabla_{\vec{R}}
    \end{align}

    Using the definitions of the Jacobi vectors in Eq. \ref{eq:jacobi_defined},
    \begin{align}
     \frac{\partial \vec{\eta}_i }{\partial \vec{r}_i } = \mathbf{0}_{3\times 3}, \quad
     \frac{\partial \vec{\lambda}_i }{\partial \vec{r}_i } = \sqrt{\frac{\mu_{i(jk)}}{m}} \mathbb{I}_3, \quad 
     \frac{\partial \vec{R}}{\partial \vec{r}_i } = \frac{m_i}{m_i + m_j+ m_k} \mathbb{I}_3
    \end{align}

    \begin{align}
     \nabla_{\vec{r}_i} = \sqrt{\frac{\mu_{i(jk)}}{m}} \nabla_{\vec{\lambda}_i} + \frac{m_i}{M} \nabla_{\vec{R}}
    \end{align}
where $M=m_i + m_j+ m_k$ is the total mass.
    \begin{align}
    \label{eq:kinetic_i}
\nabla^2_{\vec{r}_i} = \nabla_{\vec{r}_i}.\nabla_{\vec{r}_i}= \frac{\mu_{i(jk)}}{m}  \nabla^2_{\vec{\lambda}_i} + \frac{m^2_i}{M^2} \nabla^2_{\vec{R}} + t_{i3}
    \end{align}
    
where $t_{i3}$ is the off-diagonal term involving the mixed derivative. This term is given explicitly as
\begin{align}
t_{i3}=2\sqrt{\frac{\mu_{i(jk)}}{m}}\frac{m_i}{M}
\left(
\nabla_{\vec\lambda_i}\cdot\nabla_{\vec R}
\right).
\end{align}

 \item For the $j$th particle ($\nabla_{\vec{r}_j}$) \\
    \begin{align}
     \nabla_{\vec{r}_j} = \frac{\partial \vec{\eta}_i }{\partial \vec{r}_j } \nabla_{\vec{\eta}_i} + \frac{\partial \vec{\lambda}_i }{\partial \vec{r}_j } \nabla_{\vec{\lambda}_i} + \frac{\partial \vec{R}}{\partial \vec{r}_j } \nabla_{\vec{R}}
    \end{align}

    \begin{align}
     \frac{\partial \vec{\eta}_i }{\partial \vec{r}_j } = \sqrt{\frac{\mu_{jk}}{m}} \mathbb{I}_3, \quad
     \frac{\partial \vec{\lambda}_i }{\partial \vec{r}_j } = -\sqrt{\frac{\mu_{i(jk)}}{m}}\frac{m_j}{m_j + m_k} \mathbb{I}_3, \quad 
     \frac{\partial \vec{R}}{\partial \vec{r}_j } = \frac{m_j}{M}\mathbb{I}_3
    \end{align}

    \begin{align}
     \nabla_{\vec{r}_j} = \sqrt{\frac{\mu_{jk}}{m}} \nabla_{\vec{\eta}_i} -\sqrt{\frac{\mu_{i(jk)}}{m}}\frac{m_j}{m_j + m_k} \nabla_{\vec{\lambda}_i} + \frac{m_j}{M} \nabla_{\vec{R}}
    \end{align}

    \begin{align}
     \nabla^2_{\vec{r}_j} =  \nabla_{\vec{r}_j}.\nabla_{\vec{r}_j}= \frac{\mu_{jk}}{m} \nabla^2_{\vec{\eta}_i} + \frac{\mu_{i(jk)}}{m}\frac{m^2_j}{(m_j + m_k)^2}  \nabla^2_{\vec{\lambda}_i} + \frac{m^2_j}{M^2} \nabla^2_{\vec{R}} + t_{j3},
    \end{align}
    
where the sum of the off-diagonal terms $t_{j3}$ is given by 
    
\begin{align}
t_{j3}
&=
-\,2
\sqrt{\frac{\mu_{jk}}{m}}
\sqrt{\frac{\mu_{i(jk)}}{m}}
\frac{m_j}{m_j+m_k}
\;
\nabla_{\vec\eta_i}\!\cdot\!\nabla_{\vec\lambda_i}  \notag \\[6pt]
&\quad
+\,2
\sqrt{\frac{\mu_{jk}}{m}}
\frac{m_j}{M}
\;
\nabla_{\vec\eta_i}\!\cdot\!\nabla_{\vec R} \notag \\[6pt]
&\quad
-\,2
\sqrt{\frac{\mu_{i(jk)}}{m}}
\frac{m_j}{m_j+m_k}
\frac{m_j}{M}
\;
\nabla_{\vec\lambda_i}\!\cdot\!\nabla_{\vec R}
\end{align}

     \item For the $k$th particle ($\nabla_{\vec{r}_k}$) \\
    \begin{align}
     \nabla_{\vec{r}_k} = \frac{\partial \vec{\eta}_i }{\partial \vec{r}_k } \nabla_{\vec{\eta}_i} + \frac{\partial \vec{\lambda}_i }{\partial \vec{r}_k } \nabla_{\vec{\lambda}_i} + \frac{\partial \vec{R}}{\partial \vec{r}_k } \nabla_{\vec{R}}
    \end{align}

    \begin{align}
     \frac{\partial \vec{\eta}_i }{\partial \vec{r}_k } = -\sqrt{\frac{\mu_{jk}}{m}} \mathbb{I}_3, \quad
     \frac{\partial \vec{\lambda}_i }{\partial \vec{r}_k } = -\sqrt{\frac{\mu_{i(jk)}}{m}}\frac{m_k}{m_j + m_k} \mathbb{I}_3, \quad 
     \frac{\partial \vec{R}}{\partial \vec{r}_k } = \frac{m_k}{M}\mathbb{I}_3
    \end{align}

    \begin{align}
     \nabla_{\vec{r}_k} = -\sqrt{\frac{\mu_{jk}}{m}} \nabla_{\vec{\eta}_i} -\sqrt{\frac{\mu_{i(jk)}}{m}}\frac{m_k}{m_j + m_k} \nabla_{\vec{\lambda}_i} +\frac{m_k}{M} \nabla_{\vec{R}}
    \end{align}

    The Laplacian is given by
    \begin{align}
     \nabla^2_{\vec{r}_k} =  \nabla_{\vec{r}_k}.\nabla_{\vec{r}_k}= \frac{\mu_{jk}}{m} \nabla^2_{\vec{\eta}_i} + \frac{\mu_{i(jk)}}{m}\frac{m^2_k}{(m_j + m_k)^2}  \nabla^2_{\vec{\lambda}_i} + \frac{m^2_k}{M^2} \nabla^2_{\vec{R}} + t_{k3}
    \end{align}
    
The sum of the off-diagonal terms is
    
\begin{align}
t_{k3} &= 
 \quad 2 \, \frac{m_k}{m_j+m_k} \sqrt{\frac{\mu_{jk} \, \mu_{i(jk)}}{m^2}} \; (\nabla_{\vec{\eta}_i} \cdot \nabla_{\vec{\lambda}_i}) \notag \\
&\quad - \frac{2m_k}{M} \sqrt{\frac{\mu_{jk}}{m}} \; (\nabla_{\vec{\eta}_i} \cdot \nabla_{\vec{R}}) \notag \\
&\quad - \frac{2m_k^2}{M(m_j+m_k)} \sqrt{\frac{\mu_{i(jk)}}{m}} \; (\nabla_{\vec{\lambda}_i} \cdot \nabla_{\vec{R}}).
\end{align}

\end{enumerate}

The Laplacians have now been written as combinations of Laplacians in Jacobi coordinates. The next step is to substitute them into the kinetic energy operator given by Eq. \ref{eq:kinetic_single}. This substitution into the kinetic energy term is done separately for the off-diagonal and diagonal terms.

\subsubsection{Off-diagonal terms in the kinetic energy operator in Jacobi coordinates}

Substituting the off-diagonal terms into the original kinetic energy operator in Eq. \ref{eq:kinetic_single}

\begin{align}
T_3 = -\frac{\hbar^2}{2m_i}t_{i3} - \frac{\hbar^2}{2m_j}t_{j3}- \frac{\hbar^2}{2m_k}t_{k3}
\end{align}

\begin{enumerate}[(1)]
\item Terms involving $\nabla_{\vec{\eta}_i} \cdot \nabla_{\vec{\lambda}_i}$

\begin{align}
&-\frac{\hbar^2}{2m_j}\left[-2\sqrt{\frac{\mu_{jk}\mu_{i(jk)}}{m^2}} \frac{m_j}{m_j + m_k}\right] - \frac{\hbar^2}{2m_k}\left[2\sqrt{\frac{\mu_{jk}\mu_{i(jk)}}{m^2}} \frac{m_k}{m_j + m_k}\right] \nonumber \\
&= \hbar^2\sqrt{\frac{\mu_{jk}\mu_{i(jk)}}{m^2}} \left[\frac{1}{m_j + m_k} - \frac{1}{m_j + m_k}\right] \left(\nabla_{\vec{\eta}_i} \cdot \nabla_{\vec{\lambda}_i}\right) = 0
\end{align}

\item Terms involving $\nabla_{\vec{\eta}_i} \cdot \nabla_{\vec{R}}$

\begin{align}
&-\frac{\hbar^2}{2m_j}\left[2\sqrt{\frac{\mu_{jk}}{m}} \frac{m_j}{M}\right] - \frac{\hbar^2}{2m_k}\left[-2\sqrt{\frac{\mu_{jk}}{m}} \frac{m_k}{M}\right] \nonumber \\
&= -\frac{\hbar^2\sqrt{\mu_{jk}/m}}{M}\left[1 - 1\right]\left(\nabla_{\vec{\eta}_i} \cdot \nabla_{\vec{R}}\right) = 0
\end{align}

\item Terms involving $\nabla_{\vec{\lambda}_i} \cdot \nabla_{\vec{R}}$

\begin{align}
&-\frac{\hbar^2}{2m_i}\left[2\sqrt{\frac{\mu_{i(jk)}}{m}} \frac{m_i}{M}\right] - \frac{\hbar^2}{2m_j}\left[-2\sqrt{\frac{\mu_{i(jk)}}{m}} \frac{m_j^2}{(m_j + m_k)M}\right] \nonumber \\
&\quad - \frac{\hbar^2}{2m_k}\left[-2\sqrt{\frac{\mu_{i(jk)}}{m}} \frac{m_k^2}{(m_j + m_k)M}\right] \nonumber \\
&= -\frac{\hbar^2\sqrt{\mu_{i(jk)}/m}}{M}\left[1 - \frac{m_j}{m_j + m_k} - \frac{m_k}{m_j + m_k}\right]\left(\nabla_{\vec{\lambda}_i} \cdot \nabla_{\vec{R}}\right) \nonumber \\
&= -\frac{\hbar^2\sqrt{\mu_{i(jk)}/m}}{M}\left[1 - 1\right]\left(\nabla_{\vec{\lambda}_i} \cdot \nabla_{\vec{R}}\right) = 0
\end{align}

\end{enumerate}

One can immediately observe that the mass factors multiplying mixed derivatives $\nabla_{\vec{\eta}_i} \cdot \nabla_{\vec{\lambda}_i}$, $\nabla_{\vec{\eta}_i} \cdot \nabla_{\vec{R}}$ and $\nabla_{\vec{\lambda}_i} \cdot \nabla_{\vec{R}}$ are all zero. This remarkable cancellation is not coincidental: it reflects the fact that hierarchical mass scaling was specifically chosen to diagonalise the kinetic energy operator. This effectively renders the kinetic energy operator diagonal, even in cases where the mixed derivatives themselves are non-zero. 

\subsubsection{Diagonal terms in the kinetic energy operator in Jacobi coordinates}

The diagonal terms are now substituted into the original kinetic energy operator in single-particle coordinates:

\begin{align}
T = -\frac{\hbar^2}{2m_i}\nabla_{\vec{r}_i}^2 - \frac{\hbar^2}{2m_j}\nabla_{\vec{r}_j}^2 - \frac{\hbar^2}{2m_k}\nabla_{\vec{r}_k}^2
\end{align}

\begin{align}
T &= -\frac{\hbar^2}{2m_i}\left[  \frac{\mu_{i(jk)}}{m} \nabla_{\vec{\lambda}_1}^2 + \frac{m^2_i}{M^2}\nabla_{\vec{R}}^2\right] \notag \\
&- \frac{\hbar^2}{2m_j}\left[   \frac{\mu_{jk}}{m} \nabla_{\vec{\eta}_i}^2 + \frac{\mu_{i(jk)}}{m} \frac{m^2_j}{(m_j+m_k)^2} \nabla_{\vec{\lambda}_i}^2 + \frac{m^2_j}{M^2} \nabla_{\vec{R}}^2 \right] \notag \\
&- \frac{\hbar^2}{2m_k}\left[ \frac{\mu_{jk}}{m} \nabla_{\vec{\eta}_i}^2 + \frac{\mu_{i(jk)}}{m} \frac{m^2_k}{(m_j+m_k)^2} \nabla_{\vec{\lambda}_i}^2 + \frac{m^2_k}{M^2} \nabla_{\vec{R}}^2\right]
\end{align}

In order to simplify this expression for the kinetic energy, terms with the same Laplacian operator are grouped together.

\begin{enumerate}[(1)]
\item Terms with $\nabla_{\vec{\eta}_i}^2$ \\
\begin{align}
-\frac{\hbar^2}{2}\left[\frac{\mu_{jk}}{mm_j} + \frac{\mu_{jk}}{mm_k}\right]\nabla_{\vec{\eta}_i}^2 = 
-\frac{\hbar^2}{2}\left[\frac{\mu_{jk}}{m} \left(\frac{1}{m_j} + \frac{1}{m_k} \right)  \right]\nabla_{\vec{\eta}_i}^2  =  \notag \\
 -\frac{\hbar^2}{2}\left[\frac{\mu_{jk}}{m}  \frac{1}{\mu_{jk}}  \right]\nabla_{\vec{\eta}_i}^2 = 
-\frac{\hbar^2}{2m} \nabla_{\vec{\eta}_i}^2 
\end{align}

\item Terms with $\nabla_{\vec{\lambda}_i}^2$\\
\begin{align}
-\frac{\hbar^2}{2}\left[\frac{\mu_{i(jk)}}{mm_i} + \frac{\mu_{i(jk)}}{mm_j} \frac{m^2_j}{(m_j+m_k)^2} + \frac{\mu_{i(jk)}}{mm_k} \frac{m^2_k}{(m_j+m_k)^2}    \right]\nabla_{\vec{\lambda}_i}^2 =  \notag \\
-\frac{\hbar^2}{2} \frac{\mu_{i(jk)}}{m}\left[\frac{1}{m_i} + \frac{m^2_j}{m_j(m_j+m_k)^2} + \frac{m^2_k}{m_k(m_j+m_k)^2}    \right]\nabla_{\vec{\lambda}_i}^2 =  \notag \\
-\frac{\hbar^2}{2} \frac{\mu_{i(jk)}}{m}\left[\frac{1}{m_i} + \frac{1}{m_j+m_k}   \right]\nabla_{\vec{\lambda}_i}^2 = \notag  \\
-\frac{\hbar^2}{2} \frac{\mu_{i(jk)}}{m} \left[ \frac{1}{\mu_{i(jk)}}  \right] \nabla_{\vec{\lambda}_i}^2 = -\frac{\hbar^2}{2m}\nabla_{\vec{\lambda}_i}^2
\end{align} 

\item Terms with $\nabla_{\vec{R}}^2$ \\

\begin{align}
-\frac{\hbar^2}{2}\left[ \frac{1}{m_i}\frac{m^2_i}{M^2} + \frac{1}{m_j}\frac{m^2_j}{M^2} + \frac{1}{m_k}\frac{m^2_k}{M^2} \right]\nabla_{\vec{R}}^2 = \notag \\
-\frac{\hbar^2}{2M}\left[ \frac{m_i}{M} + \frac{m_j}{M} + \frac{m_k}{M} \right]\nabla_{\vec{R}}^2  
= -\frac{\hbar^2}{2M} \nabla_{\vec{R}}^2
\end{align}
\end{enumerate}
Combining all these terms, one arrives at the following: 
\begin{align}\label{eq:kinetic_jacobi}
T=  - \frac{\hbar^2}{2m}\nabla_{\vec{\eta}_i}^2 - \frac{\hbar^2}{2m}\nabla_{\vec{\lambda}_i}^2 -\frac{\hbar^2}{2M}\nabla_{\vec{R}}^2
\end{align}

This is the kinetic energy in Jacobi coordinates. The following are a few observations on the structure of this operator in Jacobi coordinates:

\begin{enumerate}[(1)]
\item The kinetic operator has a diagonalised structure since all the cross terms vanish. Furthermore, it is separated into two terms: one for the internal motion ($\nabla_{\vec{\eta}_i}^2$ and $\nabla_{\vec{\lambda}_i}^2$) and another for the centre-of-mass motion ($\nabla_{\vec{R}}^2$). This is one of the main advantages of using Jacobi coordinates in quantum mechanical three-body problems. It was shown in \cite{for2010} that coordinate systems that allow this separation into internal motion and centre-of-mass motion can be generalised to a class that is larger than Jacobi coordinates.

\item In the definition of Jacobi coordinates (Eq. \ref{eq:jacobi_defined}), a reference mass $m$ was included to scale the hierarchical mass factors. This is what resulted in the common mass factor $\hbar^2/2m$ appearing with the Laplacian operators $\nabla_{\vec{\eta}_i}^2$ and $\nabla_{\vec{\lambda}_i}^2$ in Eq. \ref{eq:kinetic_jacobi}. If this reference mass is omitted from the definition of the Jacobi coordinates, then the derivation carried out will result in a kinetic energy operator of the following form \cite{tho2009a}:

\begin{align} \label{eq:alternative}
T=  - \frac{\hbar^2}{2\mu_{jk}}\nabla_{\vec{\eta}_i}^2 - \frac{\hbar^2}{2\mu_{i(jk)}}\nabla_{\vec{\lambda}_i}^2 -\frac{\hbar^2}{2M}\nabla_{\vec{R}}^2
\end{align}
\end{enumerate}

This alternative form of the kinetic energy in Eq. \ref{eq:alternative} is also diagonalised, even though it has the disadvantage that the reduced mass factors $\hbar^2/2\mu_{jk}$ and $\hbar^2/2\mu_{i(jk)}$ would change every time the spectator in the Jacobi coordinates changes. In the case with the reference mass, the factor $\hbar^2/2m$ is fixed for all Jacobi sets. This makes computations less expensive and easier to handle. 

Although the reference mass $m$ is an arbitrary scaling parameter and does not affect the physics, it is reasonable to choose it based on the specific physical problem at hand. For example, in nuclear physics problems, $m$ is usually chosen to be the mass of a neutron (939.6 MeV/$c^2$) and in atomic physics it is the mass of an electron. In celestial mechanics, where Jacobi vectors are also used, a suitable reference mass is the mass of the Sun. For a poor choice of $m$, the mass factors multiplying the Laplacians may become extremely large or extremely small, possibly resulting in numerical difficulties such as an overflow or an underflow. As an example, consider the case of nuclear physics where the reference mass is that of a neutron with $m=939.6 MeV/c^2$. 

\begin{align}
\frac{\hbar^2}{2m} = \frac{\hbar^2c^2}{2mc^2} 
\end{align}

Since $\hbar=6.582 \times 10^{-22}$ MeV.s and $c=2.998 \times 10^{23} \,\, \text{fm}/s$, this implies $\hbar c = 197.33 \,\ \text{MeV.fm}$. Using $mc^2=939.6 MeV$,  

\begin{align}
 \frac{(\hbar c)^2}{2mc^2}  = \frac{(197.33)^2 \,\, MeV^2fm^2}{2 \times 939.6 MeV} = 20.7 MeV. fm^2
\end{align}

This is a suitable scaling parameter for most nuclear physics computations.

\section{Schr\"{o}dinger equation in Jacobi coordinates}

In Jacobi coordinates, the kinetic energy operator (Eq. \ref{eq:kinetic_jacobi}) is separated into a term for the internal dynamics of the three-body problem and a second term for the motion of the centre-of-mass. The three-body Schr\"{o}dinger equation in Jacobi coordinates may therefore be written as

\begin{align}
\label{eq:schrodinger2}
\left \{ -\frac{\hbar^2}{2m} \left(\nabla^2_{\vec{\eta}_i} + \nabla^2_{\vec{\lambda}_i} \right)  -\frac{\hbar^2}{2M} \nabla^2_{\vec{R}}  + V(\vec{\eta}_i, \vec{\lambda}_i) \right \}  \phi(\vec{\eta}_i, \vec{\lambda}_i, \vec{R})= E \phi(\vec{\eta}_i, \vec{\lambda}_i, \vec{R})
\end{align}

The potential $V(\vec{\eta}_i, \vec{\lambda}_i)$ does not depend on the centre of mass coordinate $R$ because in an isolated system the interparticle forces depend only on relative positions $(\vec{\eta}_i, \vec{\lambda}_i)$, not on the absolute position of the system in space.

The structure of the kinetic operator, coupled with the fact that the potential depends only on the internal coordinates, suggests that the total wavefunction may be decomposed as 

\begin{align}
\phi(\vec{\eta}_i, \vec{\lambda}_i, \vec{R}) = \psi(\vec{\eta}_i, \vec{\lambda}_i) \psi_{cm}(\vec{R})
\end{align}

Substituting this into Eq. \ref{eq:schrodinger2}, and writing the total as $E=E_{int} + E_{cm}$

\begin{align}
\left \{ -\frac{\hbar^2}{2m} \left(\nabla^2_{\vec{\eta}_i} + \nabla^2_{\vec{\lambda}_i} \right )  + V(\vec{\eta}_i, \vec{\lambda}_i) \right \} \psi(\vec{\eta}_i, \vec{\lambda}_i) \psi_{cm}(\vec{R}) - \frac{\hbar^2}{2M} \nabla^2_{\vec{R}}\psi(\vec{\eta}_i, \vec{\lambda}_i) \psi_{cm}(\vec{R}) = \notag \\
(E_{int} + E_{cm})\psi(\vec{\eta}_i, \vec{\lambda}_i) \psi_{cm}(\vec{R})
\end{align}

\begin{align}
\left \{ -\frac{\hbar^2}{2m} \left(\nabla^2_{\vec{\eta}_i} + \nabla^2_{\vec{\lambda}_i} \right )  + V(\vec{\eta}_i, \vec{\lambda}_i) \right \} \psi(\vec{\eta}_i, \vec{\lambda}_i) \psi_{cm}(\vec{R}) - \frac{\hbar^2}{2M} \nabla^2_{\vec{R}}\psi(\vec{\eta}_i, \vec{\lambda}_i) \psi_{cm}(\vec{R}) = \notag \\
E_{int} \psi(\vec{\eta}_i, \vec{\lambda}_i) \psi_{cm}(\vec{R}) + E_{cm} \psi(\vec{\eta}_i, \vec{\lambda}_i) \psi_{cm}(\vec{R})
\end{align}

Grouping terms

\begin{align}
&\left \{ -\frac{\hbar^2}{2m} \left(\nabla^2_{\vec{\eta}_i} + \nabla^2_{\vec{\lambda}_i} \right )  + V(\vec{\eta}_i, \vec{\lambda}_i) - E_{int} \right \} \psi(\vec{\eta}_i, \vec{\lambda}_i) \psi_{cm}(\vec{R}) \notag \\
&+ \left \{- \frac{\hbar^2}{2M} \nabla^2_{\vec{R}}   - E_{cm} \right \} \psi(\vec{\eta}_i, \vec{\lambda}_i) \psi_{cm}(\vec{R}) = 0
\end{align}

This equation can only be true for all $\vec{\eta}_i$, $\vec{\lambda}_i$ and $\vec{R}$  if

\begin{align}
\label{eq:int}
\left \{ -\frac{\hbar^2}{2m} \left(\nabla^2_{\vec{\eta}_i} + \nabla^2_{\vec{\lambda}_i} \right )  + V(\vec{\eta}_i, \vec{\lambda}_i)  \right \} \psi(\vec{\eta}_i, \vec{\lambda}_i) =  E_{int}\psi(\vec{\eta}_i, \vec{\lambda}_i) \\
\label{eq:com}
- \frac{\hbar^2}{2M} \nabla^2_{\vec{R}} \psi_{cm}(\vec{R})  =  E_{cm} \psi_{cm}(\vec{R}) 
\end{align}

Eq. \ref{eq:int} describes the internal dynamics of the three-body problem while Eq. \ref{eq:com} describes the motion of the centre-of-mass. In the absence of any external force, the centre-of-mass moves as a free particle with the total mass of the system. Therefore, it adds no information to our understanding of the three-body problem, and so it is usually ignored. 

\subsection{The centre-of-mass motion} 

The centre-of-mass equation of motion (Eq. \ref{eq:com}) describes a free particle with total mass $M$. Two types of solutions are considered: plane waves and wave packets description. The plane wave solutions are simple and pedagogical, while the wave packet description is more physically realistic.

\subsubsection{Plane-wave solution} 

The plane wave solutions are the eigenstates of the free-particle Hamiltonian. A single plane wave solution is given by

\begin{align}\label{eq:planewaves}
\psi_{cm}(\vec{R}) = N\exp{i\vec{K}.\vec{R} }
\end{align}

where $N$ is a normalisation constant and $\vec{K}$ is the wavevector associated with the centre-of-mass motion. The centre-of-mass energy is $E_{cm} = \hbar^2 K^2 / 2 M$ and linear momentum is $\vec{P}=\hbar \vec{K}$. The probability density is computed as follows:

\begin{align}
|\psi_{cm}(\vec{R})|^2 &= \psi_{cm}^*(\vec{R}) \psi_{cm}(\vec{R}) \notag \\
                       &= \left( N e^{i \vec{K} \cdot \vec{R}} \right)^* \left( N e^{i \vec{K} \cdot \vec{R}} \right)
\end{align}

Making use of the relation \((e^{i \theta})^* = e^{-i \theta}\)

\begin{align}
|\psi_{cm}(\vec{R})|^2 &= N^* e^{-i \vec{K} \cdot \vec{R}} \cdot N e^{i \vec{K} \cdot \vec{R}} \notag \\
                       &= |N|^2 e^{i \vec{K} \cdot \vec{R} - i \vec{K} \cdot \vec{R}} = |N|^2 e^{0} = N^2 
\end{align}

Therefore, the probability density of the centre of mass motion is constant i.e. it is independent of $\vec{R}$. This implies that the centre-of-mass has an equal probability of being found anywhere in space i.e., it is completely delocalised. This is an uncertainty in the position of the centre-of-mass, in the sense of Heisenberg's uncertainty principle. The plane wave solution therefore corresponds to perfect momentum definition and a complete spatial delocalisation, a scenario which is unphysical. While mathematically convenient as an eigenstate of momentum, it represents an idealisation that cannot be physically realised.

\subsubsection{Wave packet solution}

The general solution of the centre-of-mass equation is a wave packet, which is a linear superposition of plane wave solutions:
\begin{equation}
\psi_{cm}(\vec{R}) = \int A(\vec{K}) \exp(i\vec{K} \cdot \vec{R}) \, d^3K
\end{equation}
where $A(\vec{K})$ is the amplitude distribution in momentum space. This amplitude distribution characterizes the shape and localization properties of the wave packet. A common and physically relevant example is the Gaussian wave packet, where:
\begin{equation}
A(\vec{K}) = \left(\frac{2\sigma_K^2}{\pi}\right)^{3/4} \exp\left(-\sigma_K^2|\vec{K} - \vec{K}_0|^2\right)
\end{equation}

The probability density for any wave packet is computed as follows:
\begin{align}
|\psi_{cm}(\vec{R})|^2 = \psi_{cm}^*(\vec{R})\psi_{cm}(\vec{R}) 
\end{align}
\begin{align}
|\psi_{cm}(\vec{R})|^2 &= \left[\int A^*(\vec{K}') \exp(-i\vec{K}' \cdot \vec{R}) \, d^3K'\right] \left[\int A(\vec{K}) \exp(i\vec{K} \cdot \vec{R}) \, d^3K\right] \notag \\
  & = \int \int A^*(\vec{K}') A(\vec{K}) \exp(i(\vec{K} - \vec{K}') \cdot \vec{R}) \, d^3K \, d^3K' 
\end{align}
Unlike the plane wave case, this probability density is not constant, it depends on $\vec{R}$. The wave packet has a spatial extent characterised by some width $\Delta R$, and the centre of mass has a finite probability of being found in this region. This wave packet description is more realistic than the plane wave description because real systems are prepared with some localisation in space. The wave packet captures this localisation.

\subsection{Internal dynamics}
The equation describing the internal dynamics of the three-body system is

\begin{align}
\label{eq:int2}
\left \{ -\frac{\hbar^2}{2m} \left(\nabla^2_{\vec{\eta}_i} + \nabla^2_{\vec{\lambda}_i} \right )  + V(\vec{\eta}_i, \vec{\lambda}_i)  \right \} \psi(\vec{\eta}_i, \vec{\lambda}_i) =  E_{int}\psi(\vec{\eta}_i, \vec{\lambda}_i) 
\end{align}

The potential $V(\vec{\eta}_i, \vec{\lambda}_i)$ is a sum of pairwise interactions given explicitly as 

\begin{align}
V(\vec{\eta}_i, \vec{\lambda}_i) = V_{jk}(\vec{\eta}_i) + V_{ki}(\vec{\eta}_i, \vec{\lambda}_i) + V_{kj}(\vec{\eta}_i, \vec{\lambda}_i) 
\end{align}

In the Jacobi set with spectator $i$, the pairwise potential $V_{jk}$ depends only on the distance between particle j and particle k, i.e., only on the Jacobi coordinate $\vec{\eta}_i$.  In contrast, interactions involving the spectator, namely $V_{ki}$ and $V_{ij}$, generally depend on both Jacobi vectors $\vec{\eta}_i$ and $\vec{\lambda}_i$, because the distance between particle $i$ and either member of the pair $(j,k)$ is a function of both relative coordinates. Therefore, the internal dynamics of the system is described by the following Schr\"{o}dinger equation in internal Jacobi coordinates:

\begin{align}
\label{eq:int3}
\left \{ -\frac{\hbar^2}{2m} \left(\nabla^2_{\vec{\eta}_i} + \nabla^2_{\vec{\lambda}_i} \right )  + V_{jk}(\vec{\eta}_i) + V_{ki}(\vec{\eta}_i, \vec{\lambda}_i) + V_{kj}(\vec{\eta}_i, \vec{\lambda}_i)  \right \} \psi(\vec{\eta}_i, \vec{\lambda}_i) =  E_{int}\psi(\vec{\eta}_i, \vec{\lambda}_i) 
\end{align}

\subsection{Transformation between sets of Jacobi coordinates}

In many quantum-mechanical problems the wavefunction has symmetry properties that depend on the nature of the particles. This symmetry is usually encoded into the wavefunction through permutation operations on the particle coordinates. Permuting the particle coordinates is done by transforming from one set of Jacobi coordinates to another. There are three sets of Jacobi coordinates $(\vec{\eta}_1, \vec{\lambda}_1)$, $(\vec{\eta}_2, \vec{\lambda}_2)$ and $(\vec{\eta}_3, \vec{\lambda}_3)$. The transformation matrix from one set of Jacobi coordinates to another is now analysed. Consider the general transformation from $(\vec{\eta}_i, \vec{\lambda}_i)$ to $(\vec{\eta}_j, \vec{\lambda}_j)$. This transformation is in effect a rotation in the 6D space spanned by the mass-scaled Jacobi vectors. For a rotation through an angle $\theta_{ij}$ (the mixing angle) one obtains the following: 

\begin{align}
\begin{pmatrix}   
 \vec{\eta}_j  \\          
 \vec{\lambda}_j            
\end{pmatrix}  
=
G (\theta_{ij})
\begin{pmatrix}   
 \vec{\eta}_i  \\          
 \vec{\lambda}_i              
\end{pmatrix} 
\end{align}
Following the standard convention of counterclockwise rotations, the $6 \times 6$ rotation matrix is given by
\begin{align}
G = 
\begin{pmatrix}   
 \cos \theta_{ij} \mathbb{I}_3 & -\sin \theta_{ij} \mathbb{I}_3 \\          
 \sin \theta_{ij} \mathbb{I}_3 & \cos \theta_{ij} \mathbb{I}_3           
\end{pmatrix} 
\end{align}

 The mixing angle is defined for the specific transformation from the Jacobi set where particle $i$ is the spectator to the set where particle $j$ is the spectator. The matrix $G$ has a block structure since each spatial dimension transforms independently with the same $2 \times 2$ rotation. The transformation for each Cartesian component is

\begin{align}
\label{eq:cartesian_jacobi1}
\eta_{j\alpha} &= \cos\theta_{ij} \, \eta_{i\alpha} - \sin\theta_{ij} \, \lambda_{i\alpha}, \\
\label{eq:cartesian_jacobi2}
\lambda_{j\alpha} &= \sin\theta_{ij} \, \eta_{i\alpha} + \cos\theta_{ij} \, \lambda_{i\alpha},
\end{align}

where $\alpha = x, y, z$. The \( 2 \times 2 \) rotation matrix in the \(\eta\)-\(\lambda\) plane is

\begin{align}
G_{2 \times 2}(\theta_{ij}) = 
\begin{pmatrix}
\cos\theta_{ij} & -\sin\theta_{ij} \\
\sin\theta_{ij} &  \cos\theta_{ij}
\end{pmatrix},
\end{align}

This form of the rotation matrix is employed in \cite{ray1973}. In \cite{ray1970, nie2001}, an alternative form is used for the rotation matrix $G_{2 \times 2}(\theta_{ij})$. A third form for this rotation matrix is proposed in \cite{pup2009}. Using the definition  of Jacobi vectors in Eq. \ref{eq:jacobi_defined}, and through an algebraic manipulation that is not presented in this paper, it can be shown that the sine and cosine of the mixing angle are given by 

\begin{align}
\cos \theta_{ij} &= \sqrt{\frac{m_im_j}{(m_i + m_k)(m_j+m_k)}}   \\ 
\sin \theta_{ij} &= \beta_{ij} \sqrt{\frac{m_k(m_i + m_j + m_k)}{(m_i + m_k)(m_j + m_k)}}
\end{align}

where $\beta_{ij}=+1$ for a cyclic permutation of \((i,j)\) and $\beta_{ij}=-1$ for an anticyclic permutation. There are six rotations between Jacobi sets, for which $\beta_{12}=+1, \beta_{23}=+1, \beta_{31}=+1$ and $\beta_{21}=-1, \beta_{32}=-1, \beta_{13}=-1$. Of these six rotations, only three are independent since $\beta_{ij}=-\beta_{ji}$. The underlying group structure governing these rotations has been examined in \cite{sal2016}.

From the cosine and sine, the tangent is given by

\begin{align}
\tan \theta_{ij} = \beta_{ij} \sqrt{ \frac{m_k (m_i + m_j + m_k)}{m_i m_j} }.
\end{align}

The domain of the mixing angle is given by $\theta_{ij} \in [-\pi/2,\pi/2]$. This arises from the fact that all masses are positive, implying that $ \cos \theta_{ij} > 0$. This ensures that $\theta_{ij}$ comes from the fourth quadrant $ [-\pi/2,0]$ or first quadrant $ [0, \pi/2]$. Also, $\sin \theta_{ij}$ can be positive or negative, as given by the cyclic ordering. Therefore, $\sin \theta_{ij}$  determines whether the mixing angle is from the fourth quadrant or first quadrant. As an example, consider three particles with equal masses. Substituting \(m_i = m_j = m_k = m\),

\begin{align}
\cos \theta_{ij} &= \sqrt{\frac{m \cdot m}{(m + m)(m + m)}} 
                 = \frac{1}{2}, \\
\sin \theta_{ij} &= \beta_{ij} \sqrt{\frac{m(m + m + m)}{(m + m)(m + m)}}
                 = \beta_{ij} \frac{\sqrt{3}}{2}.
\end{align}

Thus $\theta_{ij} =\pm \pi/3$. The magnitude \(|\theta_{ij}| = \pi/3\) fixed, but the sign depends on the ordering of the spectator indices, reflecting whether the transformation corresponds to a cyclic (\(+\)) or anticyclic (\(-\)) permutation of the particles.

The rotation matrix G has many useful properties:

\begin{enumerate}
\item It is orthogonal i.e. $G^TG=GG^T=\mathbb{I}_6$, where $\mathbb{I}_6$ is the $6 \times 6$ identity matrix. This implies its transpose is equivalent to its inverse: $G^T=G^{-1}$. This property can be verified. Let the blocks of G be defined as

\begin{align}
A = \cos\theta_{ij} \, \mathbb{I}_3, \quad 
B = -\sin\theta_{ij} \, \mathbb{I}_3, \quad 
C = \sin\theta_{ij} \, \mathbb{I}_3, \quad 
D = \cos\theta_{ij} \, \mathbb{I}_3.
\end{align}

Then

\begin{align}
G = \begin{pmatrix} A & B \\ C & D \end{pmatrix}.
\end{align}

Since $A, B, C, D$ are scalar multiples of $\mathbb{I}_3$, they are symmetric: $A^T = A$, etc. Therefore the transpose of $G$ is given by

\begin{align}
G^T = \begin{pmatrix} A^T & C^T \\ B^T & D^T \end{pmatrix}
            = \begin{pmatrix} A & C \\ B & D \end{pmatrix}
            = \begin{pmatrix}
                \cos\theta_{ij} \, \mathbb{I}_3 & \sin\theta_{ij} \, \mathbb{I}_3 \\
                -\sin\theta_{ij} \, \mathbb{I}_3 & \cos\theta_{ij} \, \mathbb{I}_3
              \end{pmatrix}.
\end{align}

Therefore

\begin{align}
G^T G =
\begin{pmatrix} A & C \\ B & D \end{pmatrix}
\begin{pmatrix} A & B \\ C & D \end{pmatrix}
= \begin{pmatrix}
    AA + CC & AB + CD \\
    BA + DC & BB + DD
  \end{pmatrix}.
\end{align}

Computing each block:

\begin{align}
AA + CC &= (\cos^2\theta_{ij} \, \mathbb{I}_3) + (\sin^2\theta_{ij} \, \mathbb{I}_3) 
       = (\cos^2\theta_{ij} + \sin^2\theta_{ij}) \mathbb{I}_3 = \mathbb{I}_3, \\
AB + CD &= (\cos\theta_{ij})(-\sin\theta_{ij})\mathbb{I}_3 + (\sin\theta_{ij})(\cos\theta_{ij})\mathbb{I}_3 \notag \\
       &= (-\cos\theta_{ij}\sin\theta_{ij} + \sin\theta_{ij}\cos\theta_{ij}) \mathbb{I}_3 
       = \mathbf{0}_{3\times 3}, \\
BA + DC &= (-\sin\theta_{ij})(\cos\theta_{ij})\mathbb{I}_3 + (\cos\theta_{ij})(\sin\theta_{ij})\mathbb{I}_3 \notag \\
       &= (-\sin\theta_{ij}\cos\theta_{ij} + \cos\theta_{ij}\sin\theta_{ij}) \mathbb{I}_3 
       = \mathbf{0}_{3\times 3}, \\
BB + DD &= (\sin^2\theta_{ij} \, \mathbb{I}_3) + (\cos^2\theta_{ij} \, \mathbb{I}_3) 
       = (\sin^2\theta_{ij} + \cos^2\theta_{ij}) \mathbb{I}_3 = \mathbb{I}_3.
\end{align}

Assembling all the blocks, one arrives at the result

\begin{align}
G^T G =
\begin{pmatrix}
\mathbb{I}_3 & \mathbf{0}_{3\times 3} \\
\mathbf{0}_{3\times 3} & \mathbb{I}_3
\end{pmatrix}
= \mathbb{I}_6.
\end{align}

Therefore, $G^T G = \mathbb{I}_6$. It can also straightforwardly be verified that $G G^T = \mathbb{I}_6$, implying that the transformation is orthogonal. Consequently, the transformation preserves the lengths of vectors (norm conservation): $\|\vec{\eta}_j\|^2 + \|\vec{\lambda}_j\|^2 = \|\vec{\eta}_i\|^2 + \|\vec{\lambda}_i\|^2$. For the alternative transformation matrix used in \cite{ray1970, nie2001}, the determinant is also +1, revealing that it is also a pure rotation. This requirement for the transformation matrix to be orthogonal is useful in the definition of hyperspherical coordinates where the hyperradius must be invariant (preserved).
 
\item The determinant of the transformation is +1. The matrix $G$ represents the three 2D rotations applied independently to three pairs of coordinates $(x,y,z)$. Each spatial dimension contributes the same $2 \times 2$ rotation matrix $G_{2 \times 2}$. Therefore, the determinant of $G$ is given by

\begin{align}
\det(G)= [\det(G_{2 \times 2})]^3
\end{align}

It can be seen that $\det(G_{2 \times 2}) = \sin^2\theta_{ij} + \cos^2\theta_{ij} = +1$. Therefore, $\det(G)=+1$. This confirms that the transformation between Jacobi coordinate sets preserves orientation i.e. there is no reflection, it is a pure rotation. The determinant of the transformation in \cite{ray1970, nie2001} is also +1.

\item The determinant of the Jacobian matrix of the transformation is +1. The Jacobian of the transformation of $( \eta_{ix},\; \eta_{iy},\; \eta_{iz},\; \lambda_{ix},\; \lambda_{iy},\; \lambda_{iz})$ to $( \eta_{jx},\; \eta_{jy},\; \eta_{jz},\; \lambda_{jx},\; \lambda_{jy},\; \lambda_{jz})$

\begin{align}
J &= \frac{\partial ( \eta_{jx},\; \eta_{jy},\; \eta_{jz},\; \lambda_{jx},\; \lambda_{jy},\; \lambda_{jz})}{\partial (\eta_{ix},\; \eta_{iy},\; \eta_{iz},\; \lambda_{ix},\; \lambda_{iy},\; \lambda_{iz}))} 
\end{align}

From the transformation relations in Eqs. \ref{eq:cartesian_jacobi1} and \ref{eq:cartesian_jacobi2}, one may compute the partial derivatives:

\begin{align}
\frac{\partial \eta_{j\alpha}}{\partial \eta_{i\beta}} = \cos\theta_{ij} \, \delta_{\alpha\beta}, \qquad
\frac{\partial \eta_{j\alpha}}{\partial \lambda_{i\beta}} = -\sin\theta_{ij} \, \delta_{\alpha\beta},
\end{align}

\begin{align}
\frac{\partial \lambda_{j\alpha}}{\partial \eta_{i\beta}} = \sin\theta_{ij} \, \delta_{\alpha\beta}, \qquad
\frac{\partial \lambda_{j\alpha}}{\partial \lambda_{i\beta}} = \cos\theta_{ij} \, \delta_{\alpha\beta},
\end{align}

where \(\delta_{\alpha\beta}\) is the Kronecker delta. It therefore follows that the Jacobian matrix J is given by

\begin{align}
J = 
\begin{pmatrix}   
 \cos \theta_{ij} \mathbb{I}_3 & -\sin \theta_{ij} \mathbb{I}_3 \\          
 \sin \theta_{ij} \mathbb{I}_3 & \cos \theta_{ij} \mathbb{I}_3           
\end{pmatrix} 
\end{align}
This is identical to the matrix $G$, whose determinant was computed to be +1. Therefore, the determinant of the Jacobian matrix of the transformation is 1. Consequently, the transformation preserves volume elements.

\item The transformations are closed under composition: Transformations between multiple Jacobi sets (e.g. i $\to$ j $\to$ k) simplify to a single rotation by the sum of angles. This can be straightforwardly verified as follows:

\begin{align}
G(\theta_{jk}) G(\theta_{ij}) 
&= \begin{pmatrix}
\big[\cos\theta_{jk}\cos\theta_{ij} - \sin\theta_{jk}\sin\theta_{ij}\big]\mathbb{I}_3 &
-\big[\cos\theta_{jk}\sin\theta_{ij} + \sin\theta_{jk}\cos\theta_{ij}\big]\mathbb{I}_3 \notag \\
\big[\sin\theta_{jk}\cos\theta_{ij} + \cos\theta_{jk}\sin\theta_{ij}\big]\mathbb{I}_3 &
\big[\cos\theta_{jk}\cos\theta_{ij} - \sin\theta_{jk}\sin\theta_{ij}\big]\mathbb{I}_3
\end{pmatrix} \\
&= \begin{pmatrix}
\cos(\theta_{jk} + \theta_{ij}) \, \mathbb{I}_3 & -\sin(\theta_{jk} + \theta_{ij}) \, \mathbb{I}_3 \notag \\
\sin(\theta_{jk} + \theta_{ij}) \, \mathbb{I}_3 &  \cos(\theta_{jk} + \theta_{ij}) \, \mathbb{I}_3
\end{pmatrix} \\
&= G(\theta_{jk} + \theta_{ij}).
\end{align}

Therefore, $G(\theta_{jk}) G(\theta_{ij}) = G(\theta_{jk} + \theta_{ij})$. Further, it may be shown that

\begin{align}
G(\theta_{ki})G(\theta_{jk})G(\theta_{ij}) = G(\theta_{ki} + \theta_{jk} + \theta_{ij})
\end{align}

From the example of three particles with equal masses, one may observe that $\theta_{31} + \theta_{23} + \theta_{12}=\pi$ and $\theta_{13} + \theta_{32} + \theta_{21}=-\pi$. It is stated in \cite{nie2001}, and it can be easily verified, that this relation holds generally for any three different masses, with the mass scaling used in this paper.

\end{enumerate}

Having established the three-body Schr\"{o}dinger equation in Jacobi coordinates and shown how the internal and centre-of-mass motions separate, we now adopt this framework to reformulate the three-body problem in terms of Faddeev equations. These equations provide an alternative formulation of the three-body problem that exploits the Jacobi coordinate structure developed in the foregoing sections.

\section{Faddeev Equations}

\subsection{Early development of Faddeev equations}

After the discussion on Jacobi coordinates, it is now possible to properly present the Faddeev equations. These equations are an alternative to the Schr\"{o}dinger equation, for solving the three-body problem. A brief historical background to the Faddeev equations is in order. By the 1950s, the Green's function method to solve the two-body Schr\"{o}dinger equation was well-developed \cite{lip1950}. In this method,  the solution of the two-body Schr\"{o}dinger equation is written as an integral equation called the Lippmann-Schwinger equation. A seminal paper was published on the three-body problem in 1960 by Ludvig Dmitrievich Faddeev, while he was still a 27-year-old doctoral student at Steklov Mathematical Institute. This paper, which extended the Lippmann-Schwinger method from the two-body problem to the three-body problem, was written in Russian \cite{fad1960}, and later translated to English in 1961 \cite{fad1961}. Faddeev demonstrated the limitations of the Lippmann-Schwinger method for the three-body scattering problem. He went further to propose a system of coupled integral equations in momentum space for handling the three-body scattering problem. Faddeev's approach came after the failures of the extension of the M{\o}ller operator approach from the two-body problem to the three-body problem in the 1940s. His coupled integral equations are now called Faddeev equations. In Faddeev's method, the total three-body wavefunction is decomposed into a sum of three two-body wavefunctions, each associated with the scattering between a pair of particles, with the third particle being a spectator. A few months after completing his doctoral thesis, he published another paper that set the Faddeev equations on a rigorous mathematical foundation, first in Russian \cite{fad1963} and later translated into English \cite{fad1965}. These contributions by Faddeev on three-body scattering gained broader attention from the physics community only after their rigorous mathematical results were put to practical use by Lovelace on the pion-pion-nucleon system \cite{love1964}.

Faddeev's equations were originally developed in momentum space and they could only handle short-range interactions. In their original form, they face significant challenges when applied to problems involving long-range interactions, such as the Coulomb interaction between charged particles. Therefore, the tasks at hand were (i) modify the Faddeev equations to handle long-range interactions, and (ii) formulate Faddeev equations in configuration space. In configuration space, the wavefunction is expressed in terms of particle positions, while in momentum space, it is expressed in terms of particle momenta. Momentum space is a reciprocal space to configuration space, implying that the wavefunction in momentum space may be obtained by applying a Fourier transform to the wavefunction in configuration space.

The first task was pioneered through the work of Alt \textit{et al.} \cite{alt1967} and that of Noble \cite{nob1967}. The second task, namely, that of formulating the Faddeev equations in configuration space, was first achieved by Noyes and Fiedeldey \cite{noy1968}. Many more advances were made in the formulation of Faddeev equations in configuration space during the late 1960s \cite{noy1969,noy1970} and early to mid-1970s \cite{gig1972,lav1973, gig1974,mek1976}. These contributions set the Faddeev equations in configuration space on a rigorous mathematical foundation, opening the way for numerous applications.

\subsection{Faddeev equations in configuration space}

Jacobi coordinates are the natural coordinate system for Faddeev equations. The first step in writing down the Faddeev equations is to decompose the total three-body wavefunction into two-body wavefunctions, each of which depends on a set of Jacobi coordinates. 

\begin{eqnarray}
\psi = \psi_1(\vec{\eta}_1, \vec{\lambda}_1) + \psi_2(\vec{\eta}_2, \vec{\lambda}_2) + \psi_3(\vec{\eta}_3, \vec{\lambda}_3)
\end{eqnarray}

Each component $\psi_i(\vec{\eta}_i, \vec{\lambda}_i)$ satisfies a separate Faddeev equation that accounts for the interaction between the pair $(j,k)$, with $i$ being a spectator. The Faddeev equations are given by

\begin{align}
\label{eq:faddeev1}
(T_i + V_i(\vec{\eta}_i) - E)\psi_i(\vec{\eta}_i, \vec{\lambda}_i) = -V_i(\vec{\eta}_i)\left[\psi_j(\vec{\eta}_j, \vec{\lambda}_j)  +  \psi_k(\vec{\eta}_k, \vec{\lambda}_k) \right]
\end{align}

where $T_i$ is the kinetic energy operator in mass-scaled Jacobi coordinates and $V_i(\vec{\eta}_i) $ is the interaction potential between particles $j$ and $k$. As shown in the section on Jacobi coordinates, the kinetic energy operator may be written explicitly as

\begin{align}
T_i=  -\frac{\hbar^2}{2m} \left(\nabla^2_{\vec{\eta}_i} + \nabla^2_{\vec{\lambda}_i} \right )
\end{align}

Equation \ref{eq:faddeev1} can be written explicitly as follows:

\begin{subequations}\label{eq:faddeev2}
\begin{align}
(T_1 + V_1(\vec{\eta}_1) - E)\psi_1(\vec{\eta}_1, \vec{\lambda}_1) = -V_1(\vec{\eta}_1)\left[\psi_2(\vec{\eta}_2, \vec{\lambda}_2)  +  \psi_3(\vec{\eta}_3, \vec{\lambda}_3) \right] \\
(T_2 + V_2(\vec{\eta}_2) - E)\psi_2(\vec{\eta}_2, \vec{\lambda}_2) = -V_2(\vec{\eta}_2)\left[\psi_3(\vec{\eta}_3, \vec{\lambda}_3)  +  \psi_1(\vec{\eta}_1, \vec{\lambda}_1) \right] \\
(T_3 + V_3(\vec{\eta}_3) - E)\psi_3(\vec{\eta}_3, \vec{\lambda}_3) = -V_3(\vec{\eta}_3)\left[\psi_1(\vec{\eta}_1, \vec{\lambda}_1)  +  \psi_2(\vec{\eta}_2, \vec{\lambda}_2) \right]
\end{align}
\end{subequations}
This is a system of coupled second-order differential equations. It should be emphasised that the equations are written in spectator notation:

\begin{enumerate}
    \item $V_1(\vec{\eta}_1)$ describes the interaction between particles (2,3) while particle 1 is a spectator. Also $\psi_1(\vec{\eta}_1, \vec{\lambda}_1)$ is the wavefunction of the (2,3) pair while 1 is a spectator.
    \item $V_2(\vec{\eta}_2)$ describes the interaction between particles (3,1) while particle 2 is a spectator. Also $\psi_2(\vec{\eta}_2, \vec{\lambda}_2)$ is the wavefunction of the (3,1) pair while 2 is a spectator.
    \item $V_3(\vec{\eta}_3)$ describes the interaction between particles (1,2) while particle 3 is a spectator. Also $\psi_3(\vec{\eta}_3, \vec{\lambda}_3)$ is the wavefunction of the (1,2) pair while 3 is a spectator.
\end{enumerate}

Faddeev's equations are preferable to the three-body Schr\"{o}dinger equation for describing the three-body problem. Some of the reasons are discussed here.

\begin{enumerate}[(1)]
    \item Avoiding multi-particle asymptotics in boundary conditions \\
    
    Boundary conditions for the quantum-mechanical problem are usually stated as asymptotic conditions on the wavefunction: this is the behaviour of the wavefunction at large distances (i.e., when particles are far apart). In the three-body Schr\"{o}dinger equation it is very difficult to set boundary conditions for the scattering problem due to the fact that the asymptotic conditions on the wavefunction must account for the possibility of different rearrangement channels, for example one body scattering off a bound state of the other two. Since Faddeev's equations only deal with the interactions between two bodies at a time, it is easier to state the boundary conditions. In the case of a three-body breakup problem where all three particles separate, Schr\"{o}dinger approach often struggles to converge in the presence of this complicated multi-particle asymptotics. \\

    \item Avoiding the overcounting problem \\

    In the three-body Schr\"{o}dinger equation, the sum of pairwise interactions acts on the full three-body wavefunction. This causes the dynamical influence of each pair to be entangled with the others from the outset. This entanglement complicates the interpretation of two-body correlations and can lead to redundant dynamical coupling in iterative or perturbative treatments, a situation sometimes referred to as an ``overcounting'' problem.
    
In contrast, the Faddeev equations isolate each pair interaction into a separate component equation. The potential $V_i$ acts only on its corresponding component $\psi_i$, while the other components $\psi_j$ and $\psi_k$ enter as source terms. This structure cleanly separates the two-body dynamics and avoids the ambiguous coupling present in the Schrödinger formulation, thereby eliminating the overcounting issue. \\

    \item Isolating singularities in potentials \\

    Many nuclear potentials have a repulsive core which may be hard or soft. When two bodies get too close to each other, this repulsive core causes the potentials to display a divergent behaviour: an infinite repulsion for the hard-core potentials or a steep but finite divergence for the soft-core potentials. This situation may lead to non-convergent or ill-defined solutions. In the Faddeev equations this singularity is isolated and regularisation is done locally within the affected equation, whereas in the three-body Schr\"{o}dinger equation the singularity within a single interacting pair affects the entire three-body problem. Since soft core potentials are more realistic, it is not surprising that the modern potentials used in nuclear physics are soft-core e.g. Argonne v18 potential. 
\end{enumerate}

\subsection{Two indistinguishable particles among three particles}

In many three-body problems, two particles are indistinguishable. For example, the triton (p + n + n), helium-3 nucleus (n + p + p), beryllium-9 ($\alpha$ + $\alpha$ + n), boron-11 (3He + $\alpha$ + $\alpha$) and lithium-11 (9Li + n + n). When two particles are indistinguishable, this imposes symmetry or antisymmetry constraints on the two-body wavefunctions. If the pair of indistinguishable particles are bosons, then symmetry constraints are enforced. For a pair of indistinguishable fermions, there is antisymmetry in the wavefunction. Let us assume, without any loss of generality, that particles 2 and 3 are indistinguishable.  Furthermore, let $\mathbf{P}_{23}$ be the permutation operator that acts on two-body wavefunctions to interchange coordinates of particles 2 and 3. This operator has the following actions:

\begin{subequations} \label{eq:permutations}
\begin{align}
\mathbf{P}_{23} \psi_1(\vec{\eta}_1, \vec{\lambda}_1)&= \psi_1(\vec{\eta}_1, \vec{\lambda}_1) \\
\mathbf{P}_{23} \psi_2(\vec{\eta}_2, \vec{\lambda}_2)&= \psi_3(\vec{\eta}_3, \vec{\lambda}_3)=p_{23} \psi_2(\vec{\eta}_2, \vec{\lambda}_2) \\
\mathbf{P}_{23} \psi_3(\vec{\eta}_3, \vec{\lambda}_3)&= \psi_2(\vec{\eta}_2, \vec{\lambda}_2)=p_{23} \psi_3(\vec{\eta}_3, \vec{\lambda}_3)
\end{align}
\end{subequations}

where $p_{23}$ is the eigenvalue of the operator $\mathbf{P}_{23}$ with values $p_{23}=+1$ if particles 2 and 3 are bosons and $p_{23}=-1$ if they are fermions. Since the two-body wavefunction $\psi_1(\vec{\eta}_1, \vec{\lambda}_1)$ describes the partition in which the (2,3) pair is interacting, the operator $\mathbf{P}_{23}$ leaves this wavefunction unchanged. The operator $\mathbf{P}_{23}$ has three useful properties:
\begin{enumerate}
\item It is hermitian: $\mathbf{P}^{\dagger}_{23} = \mathbf{P}_{23}$ 
\item It is unitary: $\mathbf{P}^{\dagger}_{23}\mathbf{P}_{23} = I$, where I is the identity operator
\item Involution: Applying the permutation twice restores the original wavefunction: $\mathbf{P}^{2}_{23} = I$, where I is the identity operator. 
\end{enumerate}

Since particles 2 and 3 are indistinguishable, one may choose to write the Faddeev equations in terms of $\psi_1(\vec{\eta}_1, \vec{\lambda}_1)$ and $\psi_2(\vec{\eta}_2, \vec{\lambda}_2)$, eliminating $\psi_3(\vec{\eta}_3, \vec{\lambda}_3)$. Alternatively, one may write the Faddeev equations in terms of $\psi_1(\vec{\eta}_1, \vec{\lambda}_1)$ and $\psi_3(\vec{\eta}_3, \vec{\lambda}_3)$, eliminating $\psi_2(\vec{\eta}_2, \vec{\lambda}_2)$. In this survey, the former approach is used. 

In the original Faddeev equations in Eq. \ref{eq:faddeev2} we use the permutations in Eq. \ref{eq:permutations} to eliminate $\psi_3(\vec{\eta}_3, \vec{\lambda}_3)$. Since particles 2 and 3 are indistinguishable, the permutation symmetry gives

\begin{align}
\psi_3(\vec{\eta}_3, \vec{\lambda}_3) = p_{23} \psi_2(\vec{\eta}_2, \vec{\lambda}_2),
\end{align}

and the potentials satisfy \(V_2(\vec{\eta}_2) = V_3(\vec{\eta}_3)\) because the interaction between particles (3,1) is the same as between (1,2) when particles 2 and 3 are identical. The procedure to reduce the Faddeev equations is carried out in three steps.

\begin{enumerate}[(1)]
\item Step 1: Substitute \(\psi_3 = p_{23} \psi_2\) into the first Faddeev equation:

\begin{align}
\label{eq:step1}
(T_1 + V_1 - E) \psi_1 = -V_1 \bigl[ \psi_2 + p_{23} \psi_2 \bigr] = -V_1 (1 + p_{23}) \psi_2.
\end{align}

\item Step 2: Substitute \(\psi_3 = p_{23} \psi_2\) into the second Faddeev equation:

\begin{align}
\label{eq:step2}
(T_2 + V_2 - E) \psi_2 = -V_2 \bigl[ p_{23} \psi_2 + \psi_1 \bigr].
\end{align}

\item Step 3: In this step, it is shown that the third Faddeev equation is redundant.  
Substituting \(\psi_3 = p_{23} \psi_2\) and \(V_3 = V_2\) into the third equation gives:

\begin{align}
(T_2 + V_2 - E)(p_{23} \psi_2) = -V_2 \bigl[ \psi_1 + \psi_2 \bigr].
\end{align}

Since \(T_2\) and \(V_2\) commute with the constant \(p_{23}\), this simplifies to:

\begin{align}
p_{23} (T_2 + V_2 - E) \psi_2 = -V_2 \bigl[ \psi_1 + \psi_2 \bigr].
\end{align}

Multiplying Eq.\ref{eq:step2} by \(p_{23}\) yields:

\begin{align}
p_{23} (T_2 + V_2 - E) \psi_2 = -V_2 \bigl[ p_{23}^2 \psi_2 + p_{23} \psi_1 \bigr] = -V_2 \bigl[ \psi_2 + p_{23} \psi_1 \bigr],
\end{align}

since \(p_{23}^2 = 1\). Consistency with the third equation requires \(p_{23} \psi_1 = \psi_1\), which holds because \(\psi_1\) is symmetric under exchange of particles 2 and 3. Thus, the third equation does not provide new information; it is automatically satisfied if Eqs.~\ref{eq:step1} and \ref{eq:step2} hold.
\end{enumerate}

The system therefore reduces to two coupled Faddeev equations:

\begin{subequations} \label{eq:faddeev_reduced}
\begin{align}
(T_1 + V_1(\vec{\eta}_1) - E)\psi_1(\vec{\eta}_1, \vec{\lambda}_1) &= -V_1(\vec{\eta}_1)\left[\psi_2(\vec{\eta}_2, \vec{\lambda}_2)  +  p_{23} \psi_2(\vec{\eta}_2, \vec{\lambda}_2) \right], \\
(T_2 + V_2(\vec{\eta}_2) - E) \psi_2(\vec{\eta}_2, \vec{\lambda}_2) &= -V_2(\vec{\eta}_2)\left[   p_{23} \psi_2(\vec{\eta}_2, \vec{\lambda}_2) + \psi_1(\vec{\eta}_1, \vec{\lambda}_1)  \right],
\end{align}
\end{subequations}

or more compactly:

\begin{align}
(T_1 + V_1 - E) \psi_1 &= -V_1 (1 + p_{23}) \psi_2, \\
(T_2 + V_2 - E) \psi_2 &= -V_2 \bigl( p_{23} \psi_2 + \psi_1 \bigr).
\end{align}

This system, together with the symmetry condition \(\psi_3(\vec{\eta}_3, \vec{\lambda}_3) = p_{23} \psi_2(\vec{\eta}_2, \vec{\lambda}_2)\), fully defines the three-body problem whenever two particles are indistinguishable. The total wavefunction is given by:

\begin{align}
\psi = \psi_1(\vec{\eta}_1, \vec{\lambda}_1) + \psi_2(\vec{\eta}_2, \vec{\lambda}_2) + \psi_3(\vec{\eta}_3, \vec{\lambda}_3) = \psi_1(\vec{\eta}_1, \vec{\lambda}_1) + (1 + p_{23})\psi_2(\vec{\eta}_2, \vec{\lambda}_2).
\end{align}

\subsection{Three indistinguishable particles}

In some cases, a three-body problem has all three particles being indistinguishable particles. For example carbon-12 ($\alpha$ + $\alpha$ +$\alpha$). The three-body permutation operator $\mathbf{P}$, called a cyclic permutation operator, is introduced. This operator carries out the following action: 1 $\to$ 2, 2 $\to$ 3 and 3 $\to$ 1 i.e. (1,2,3) changes to (2,3,1). The anticyclic permutation operator, which is the inverse of the cyclic permutation operator is $\mathbf{P}^{-1}$. This inverse operator carries out the following action: 1 $\to$ 3, 2 $\to$ 1 and 3 $\to$ 2 i.e. (1,2,3) changes to (3, 1, 2). The cyclic and anticyclic permutation operators can be expressed in terms of two-body permutation operators as follows: $\mathbf{P} =  \mathbf{P}_{12} \mathbf{P}_{23}$ and $\mathbf{P}^{-1} =\mathbf{P}_{23} \mathbf{P}_{12}$. It can easily be shown that $\mathbf{P}^3=I$, unlike the two-body permutation operators that have the property of involution. This means applying $\mathbf{P}$ three times restores the original order of the particles. Also $\mathbf{P}^2=\mathbf{P}^{-1} $ i.e. applying the cyclic permutation twice is equivalent to applying one anticyclic permutation.

The operators $\mathbf{P}$ and $\mathbf{P}^{-1}$ carry out the following actions: 

\begin{subequations} \label{eq:permutations2}
\begin{align}
\mathbf{P} \psi_1(\vec{\eta}_1, \vec{\lambda}_1)&= \psi_2(\vec{\eta}_2, \vec{\lambda}_2) = p \psi_1(\vec{\eta}_1, \vec{\lambda}_1)  \\
\mathbf{P}^{-1} \psi_1(\vec{\eta}_1, \vec{\lambda}_1)&= \psi_3(\vec{\eta}_3, \vec{\lambda}_3) =p^{-1} \psi_1(\vec{\eta}_1, \vec{\lambda}_1)
\end{align}
\end{subequations}

where $p$ and $p^{-1}$ are the eigenvalues of the operators ($p^{-1}$ should not be mistaken for $1/p$). Since all three particles are indistinguishable, the potentials are the same in all partitions i.e. $V_1(\vec{\eta}_1)=V_2(\vec{\eta}_2) = V_3(\vec{\eta}_3)=V$. The wavefunctions $\psi_2(\vec{\eta}_2, \vec{\lambda}_2)$  and $\psi_3(\vec{\eta}_3, \vec{\lambda}_3)$ are now eliminated from the original Faddeev equations in Eq. \ref{eq:faddeev2}. Since all three particles are indistinguishable, we have the symmetry relations from Eq.~\ref{eq:permutations2}: $\psi_2 = p\psi_1$ and $\psi_3 = p^{-1}\psi_1$. As in the previous section, the procedure to reduce the Faddeev equations is carried out in three steps.

\begin{enumerate}[(1)]

\item Step 1: Substituting \(\psi_2 = p\psi_1\) and \(\psi_3 = p^{-1}\psi_1\) into the first Faddeev equation in Eq. \ref{eq:faddeev2}:

\begin{align}\label{eq:step1_three}
(T_1 + V - E) \psi_1 = -V \bigl[ p\psi_1 + p^{-1}\psi_1 \bigr] = -V(p + p^{-1}) \psi_1.
\end{align}

\item Step 2: This step seeks to verify that the second and third Faddeev equations are automatically satisfied and provide no new information. Starting with the second equation \ref{eq:faddeev2}:

\begin{align}
(T_2 + V - E) \psi_2 = -V \bigl[ \psi_3 + \psi_1 \bigr].
\end{align}

Substituting \(\psi_2 = p\psi_1\) and \(\psi_3 = p^{-1}\psi_1\):

\begin{align}
(T_2 + V - E)(p\psi_1) = -V \bigl[ p^{-1}\psi_1 + \psi_1 \bigr].
\end{align}

Since the kinetic energy operator \(T_2\) is identical in form to \(T_1\), and all particles are identical, we have \(T_2\psi_1 = T_1\psi_1\). Therefore:

\begin{align}
p(T_1 + V - E)\psi_1 = -V(1 + p^{-1})\psi_1.
\end{align}

From Eq.~\ref{eq:step1_three}, we know \((T_1 + V - E)\psi_1 = -V(p + p^{-1})\psi_1\). Substituting this:

\begin{align}
p[-V(p + p^{-1})\psi_1] = -V(1 + p^{-1})\psi_1.
\end{align}

Simplifying both sides:

\begin{align}
-V(p^2 + pp^{-1})\psi_1 = -V(1 + p^{-1})\psi_1 \\
-V(p^2 + 1)\psi_1 = -V(1 + p^{-1})\psi_1 
\end{align}

Since $\mathbf{P}^2=\mathbf{P}^{-1} $, it follows that the eigenvalues are related as \(p^{-1} = p^2\). Therefore,

\begin{align}
-V(p^2 + 1)\psi_1 = -V(1 + p^2)\psi_1,
\end{align}

which is identically satisfied. The same verification holds for the third equation.
\end{enumerate}

Rearrange Eq.~\ref{eq:step1_three} to obtain the final single Faddeev equation. Starting from:

\begin{align}
(T_1 + V - E) \psi_1 = -V(p + p^{-1}) \psi_1,
\end{align}

we bring all terms involving \(\psi_1\) to the left-hand side:

\begin{align}
(T_1 + V - E) \psi_1 + V(p + p^{-1}) \psi_1 = 0.
\end{align}

Factor out \(\psi_1\):

\begin{align}
[T_1 + V(1 + p + p^{-1}) - E] \psi_1 = 0.
\end{align}

Since \(1 + p + p^{-1}\) is a constant factor, we can write:

\begin{align}
(T_1 - E)\psi_1 = -V(1 + p + p^{-1})\psi_1.
\end{align}

The three Faddeev equations thus reduce to a single equation,

\begin{align}
(T_1 + V(\vec{\eta}_1) - E)\psi_1(\vec{\eta}_1,\vec{\lambda}_1) = -V(\vec{\eta}_1)\left[p\psi_1(\vec{\eta}_1,\vec{\lambda}_1) + p^{-1}\psi_1(\vec{\eta}_1,\vec{\lambda}_1)\right]
\end{align}

or equivalently:

\begin{align}
(T_1 - E)\psi_1(\vec{\eta}_1,\vec{\lambda}_1) = -V(\vec{\eta}_1)\left[1 + p + p^{-1}\right]\psi_1(\vec{\eta}_1,\vec{\lambda}_1)
\end{align}

The total wavefunction is

\begin{align}
\psi &= \psi_1(\vec{\eta}_1,\vec{\lambda}_1) + \psi_2(\vec{\eta}_2,\vec{\lambda}_2) + \psi_3(\vec{\eta}_3,\vec{\lambda}_3) \notag \\
     &= \psi_1(\vec{\eta}_1,\vec{\lambda}_1) + p\psi_1(\vec{\eta}_1,\vec{\lambda}_1) + p^{-1}\psi_1(\vec{\eta}_1,\vec{\lambda}_1) = (1 + p + p^{-1})\psi_1(\vec{\eta}_1,\vec{\lambda}_1).
\end{align}

This single equation, along with the cyclic symmetry conditions (wavefunction relations, potential equality, and eigenvalue properties), fully defines the three-body problem when all three particles are indistinguishable.

\section{Faddeev equations in hyperspherical coordinates}

As discussed earlier, the main advantage of Jacobi coordinates is that the kinetic energy operator becomes separated into terms for the internal relative motion and the centre-of-mass motion. For an isolated system, the centre-of-mass motion is ignored, thereby reducing the dimensions of the problem from 9 to 6. In addition to this, Jacobi coordinates have other advantages for specific type of problems. For example, they are ideal for a hierarchical system i.e a three-body problem in which there is a two-body bound state and a third distant body, such a deuteron plus a third particle. 

In spite of these advantages, the three-body problem is now more frequently formulated in terms of hyperspherical coordinates than Jacobi coordinates. There are good reasons for this, some of which will become evident. Faddeev equations in hyperspherical coordinates or the three-body Schr\"{o}dinger equation in hyperspherical coordinates offer more advantages compared to the case where these equations are written in Jacobi coordinates.

\subsection{Delves hyperspherical coordinates}

Hyperspherical coordinates generalize the familiar spherical coordinates to Euclidean spaces of dimension greater than three. In such systems, a hyperradius describes the overall scale of the system, while angular coordinates parametrise the surface of a higher‑dimensional sphere called a hypersphere.

The mathematical foundation for higher-dimensional spheres began in the mid-19th century with Bernhard Riemann's work on manifolds and Ludwig Schläfli's studies of polytopes and spherical geometry in $n$ dimensions. By the late 19th and early 20th centuries, the term hyperspherical had entered widespread mathematical usage, notably through contributions by Hermann von Helmholtz on spaces of constant curvature, Élie Cartan on symmetric spaces and harmonic analysis, Hermann Weyl on group-theoretic methods in higher-dimensional spaces, and Henri Poincaré on topology and automorphic forms.

In physics, hyperspherical coordinates were first applied to quantum systems in the 1930s. In 1935 Vladimir Fock used a four‑dimensional hypersphere in momentum space to solve the hydrogen atom \cite{foc1935}, while in 1937 T. H. Gronwall introduced a related coordinate system for the helium atom \cite{gro1937} (posthumously published). Gronwall introduced a coordinate transformation for the helium atom that can be viewed as an early precursor to hyperspherical ideas (introducing a hyperradius-like variable and related angles in a modified spherical framework for two-electron systems). While not identical to modern hyperspherical coordinates, it is commonly cited in historical reviews as an important early step toward hyperspherical treatments of helium \cite{lin1995}. The hyperspherical method was introduced in nuclear physics by L. M. Delves in 1958 through his treatment of the triton \cite{del1958}, paving the way for its widespread adoption in few‑body quantum mechanics. During the period from the 1960s to the 1980s numerous contributions were made by J.H. Ma\v{c}ek, J.S. Avery, Yu. A. Simonov, M. Fabre de la Ripelle, etc. to establish hyperspherical coordinates into the useful tool that it is today in nuclear physics. 

Hyperspherical surfaces generalize circles (1-spheres) and ordinary spheres (2-spheres) to dimensions of four and higher. Generally, an $n$-sphere $S^{n}$ is the set of points (a hypersurface) in an ($n+1$)-dimensional Euclidean space ($\mathbb{R}^{n+1}$) at a fixed distance from the origin. For example, a 1-sphere (circle) is a hypersurface in the 2-dimensional space ($\mathbb{R}^{2}$), a 2-sphere (an ordinary sphere) is a hypersurface in the 3-dimensional space ($\mathbb{R}^{3}$), a 3-sphere is a hypersurface in a 4-dimensional space ($\mathbb{R}^{4}$), a 4-sphere is a hypersurface in the 5-dimensional space ($\mathbb{R}^{5}$) and a 5-sphere is a hypersurface in the 6-dimensional space ($\mathbb{R}^{6}$). In six-dimensional Euclidean space $\mathbb{R}^6$, hyperspherical coordinates consist of one hyperradius and five angles. These angles parametrise the surface of the 5-sphere $S^5$, which corresponds to points at a fixed distance from the origin. This fixed distance is called the hyperradius.

Two of the most commonly used hyperspherical coordinates in nuclear physics are Delves coordinates and Smith-Whitten coordinates. From the six-dimensional Jacobi coordinates $(\vec{\eta}_i, \vec{\lambda}_i)=(\eta_{ix}, \eta_{iy}, \eta_{iz}, \lambda_{ix}, \lambda_{iy}, \lambda_{iz})$, six-dimensional hyperspherical coordinates ($\rho$, $\theta_i$, $\nu_{\eta_i}$,  $\omega_{\eta_i}$, $\nu_{\lambda_i}$, $\omega_{\lambda_i}$) are constructed. For a three-body problem, Delves hyperspherical coordinates consist of one hyperradius, one hyperangle and four polar angles. Smith-Whitten hyperspherical coordinates (also called kinematic or body-fixed hyperspherical coordinates) in six dimensions consists of one hyperradius, two hyperangles and three Euler angles. Smith-Whitten coordinates are not used in this survey. Delves coordinates are defined as follows:

\begin{enumerate}[(1)]
\item The coordinate $\rho \in [0, \infty)$ is called the hyperradius and it is defined as follows:

\begin{align}
\rho = \sqrt{\eta^2_i + \lambda^2_i}
\end{align}

where $\eta_i$ and $\lambda_i$ are the previously Jacobi coordinates. This is a collective coordinate that defines the overall size of the system. For example, when the system breaks up, $\rho$ goes to infinity. The hyperradius does not change when the Jacobi set changes i.e. it is invariant under rotations in the six-dimensional space. Therefore,

\begin{align}
\rho = \sqrt{\eta^2_i + \lambda^2_i} = \sqrt{\eta^2_j + \lambda^2_j} = \sqrt{\eta^2_k + \lambda^2_k}
\end{align}

\item The coordinate $\theta_i \in [0, \pi/2]$ is called the hyperangle. It is defined through the Cartesian parametrisation 
\begin{subequations} \label{eq:delves}
\begin{align}
\eta_i= \rho \sin \theta_i \\
\lambda_i = \rho \cos \theta_i
\end{align}

from which 

\begin{align}
\theta_i = \arctan \left ( \frac{\eta_i}{\lambda_i}\right)
\end{align}
\end{subequations}

\item The pair of coordinates $(\nu_{\eta_i},  \omega_{\eta_i})$ are spherical polar angles associated with the Jacobi vector $\vec{\eta}_i$ while $(\nu_{\lambda_i}, \omega_{\lambda_i})$ are spherical polar angles associated with $\vec{\lambda}_i$. This means $(\eta_i, \nu_{\eta_i},  \omega_{\eta_i})$ and $(\lambda_i, \nu_{\lambda_i}, \omega_{\lambda_i})$ form separate spherical coordinate systems. While the hyperradius defines the overall size of the system, these angular coordinates describe the internal structure.  For $\vec{\eta}_i$

\begin{align} \label{eq:spherical1}
\eta_{ix} &= \eta_i \sin\nu_{\eta_i} \cos\omega_{\eta_i} \\
\eta_{iy} &= \eta_i \sin\nu_{\eta_i} \sin\omega_{\eta_i} \\
\eta_{iz} &= \eta_i \cos\nu_{\eta_i}
\end{align}

and for $\vec{\lambda}_i$

\begin{align}\label{eq:spherical2}
\lambda_{ix} &= \lambda_i \sin\nu_{\lambda_i} \cos\omega_{\lambda_i} \\
\lambda_{iy} &= \lambda_i \sin\nu_{\lambda_i} \sin\omega_{\lambda_i} \\
\lambda_{iz} &= \lambda_i \cos\nu_{\lambda_i}
\end{align}
\end{enumerate}
  
The Jacobian determinant of the transformation from $(\vec{\eta}_i, \vec{\lambda}_i)=(\eta_{ix}, \eta_{iy}, \eta_{iz}, \lambda_{ix}, \lambda_{iy}, \lambda_{iz})$ to $(\rho, \theta_i, \nu_{\eta_i},  \omega_{\eta_i}, \nu_{\lambda_i}, \omega_{\lambda_i})$ is now calculated. The Jacobian matrix, $J$ has dimensions of $6 \times 6$, and its determinant is given by

\begin{align}
\det( J)=\left| \frac{\partial( \rho, \theta_i, \nu_{\eta_i}, \omega_{\eta_i}, \nu_{\lambda_i}, \omega_{\lambda_i} )}{\partial( \eta_{ix}, \eta_{iy}, \eta_{iz}, \lambda_{ix}, \lambda_{iy}, \lambda_{iz} )} \right|
\end{align}

It is a tedious exercise to compute this determinant directly from the partial derivatives. A simpler and more elegant approach is to exploit the structure of the transformation: since $\vec{\eta_i}$ and $\vec{\lambda_i}$ transform independently, their transformations form blocks in the Jacobian matrix, which allows it to be decomposed into smaller sequential transformations. First, there are two transformations from Cartesian to spherical coordinates, and then there is the magnitude transformation from $(\eta_i, \lambda_i)$ to $(\rho, \theta_i)$. These three sequential transformations are as follows

\begin{enumerate}[(1)]
    \item Step 1: Transformation $(\eta_{ix}, \eta_{iy}, \eta_{iz}) \to (\eta_i, \nu_{\eta_i}, \omega_{\eta_i})$, with Jacobian $J_A$.
    \item Step 1: Transformation $(\lambda_{ix}, \lambda_{iy}, \lambda_{iz}) \to (\lambda_i, \nu_{\lambda_i}, \omega_{\lambda_i})$, with Jacobian $J_B$.
    \item Step 3: Transformation of the the magnitudes $(\eta_i, \lambda_i) \to (\rho, \theta_i)$, with Jacobian $J_C$.
\end{enumerate}

The determinant of the full Jacobian is therefore given by

\[
\det(J_{\text{total}})= 
\left|\frac{\partial( \eta_i,\nu_{\eta_i},\omega_{\eta_i} )}
{\partial( \eta_{ix},\eta_{iy},\eta_{iz}     )}\right|
\;\times\;
\left|\frac{\partial(\lambda_i,\nu_{\lambda_i},\omega_{\lambda_i}      )}
{\partial(  \lambda_{ix},\lambda_{iy},\lambda_{iz}  )}\right|
\;\times\;
\left|\frac{\partial(  \rho,\theta_i)}{\partial( \eta_i,\lambda_i  )}\right|
\]

The Jacobian determinant for each of these steps is given as follows:

\begin{enumerate}[(1)]

\item Step 1: It is a known result that the Jacobian determinant of the transformation from Cartesian to standard spherical coordinates is $r^2\sin\theta$, $r$ is the radius. Therefore, transformation from $(\eta_{ix}, \eta_{iy}, \eta_{iz})$ to $(\eta_i, \nu_{\eta_i}, \omega_{\eta_i})$ is given by

\begin{align}
|\det(J_A)| = \eta_i^2 \sin\nu_{\eta_i}
\end{align}

\item Step 2: In like manner as in Step 1, the transformation from $(\lambda_{ix}, \lambda_{iy}, \lambda_{iz})$ to $(\lambda_i, \nu_{\lambda_i}, \omega_{\lambda_i})$ has a Jacobian determinant given by

\begin{align}
|\det(J_B)| = \lambda_i^2 \sin\nu_{\lambda_i}
\end{align}

\item Step 3: The Jacobian determinant for the transformation of magnitudes $(\eta_i, \lambda_i) \to (\rho, \theta_i)$ is given by 

\begin{align}
\det( J_C)&=\left| \frac{\partial(\eta_i, \lambda_i)}{\partial(\rho, \theta_i)} \right| 
 = \begin{vmatrix}
\sin \theta_i & \rho \cos \theta_i \notag \\
\cos \theta_i & -\rho \sin \theta_i
\end{vmatrix} \\
&= -\rho \sin^2 \theta_i - \rho \cos^2 \theta_i = -\rho  \\
|\det( J_C)|&= \rho
\end{align}
\end{enumerate}

The full Jacobian determinant is therefore given by

\begin{align}
|\det(J)| &= |\det(J_A)| \times | \det(J_B)| \times |\det(J_B)| \notag  \\
          &=  \eta_i^2 \sin\nu_{\eta_i} \times \lambda_i^2 \sin\nu_{\lambda_i} \times \rho   
\end{align}

Making use of the parametrisations $\eta_i= \rho \sin \theta_i$ and $\lambda_i = \rho \cos \theta_i$, this determinant is given explicitly as

\begin{align}
|\det(J)|  &=  \rho^2\sin^2\theta_i \sin\nu_{\eta_i} \times \rho^2\cos^2\theta_i \sin\nu_{\lambda_i} \times \rho  \notag \\
        &= \rho^5 \sin^2\theta_i \cos^2\theta_i \sin\nu_{\eta_i} \sin\nu_{\lambda_i}  
\end{align}

\subsection{Volume element} 

From this Jacobian determinant, it follows that volume elements transform as follows: 

\begin{align}
 d\eta_{ix} d\eta_{iy} d\eta_{iz} d\lambda_{ix} d\lambda_{iy} d\lambda_{iz} = \rho^5 \sin^2\theta_i \cos^2\theta_i \sin\nu_{\eta_i} \sin\nu_{\lambda_i} d\rho d\theta_i d\nu_{\eta_i} d\omega_{\eta_i} d\nu_{\lambda_i} d\omega_{\lambda_i}   
\end{align}

To simplify the notation, the five angles are collectively represented as $d \Omega_5 = d\theta_i d\nu_{\eta_i} d\omega_{\eta_i} d\nu_{\lambda_i} d\omega_{\lambda_i}$ and $d^3\vec{\eta}_i\, d^3\vec{\lambda}_i = d\eta_{ix} d\eta_{iy} d\eta_{iz} d\lambda_{ix} d\lambda_{iy} d\lambda_{iz}$. The volume element transformation is written as

\begin{align}
 d^3\vec{\eta}_i\, d^3\vec{\lambda}_i = \rho^5 \sin^2\theta_i \cos^2\theta_i \sin\nu_{\eta_i} \sin\nu_{\lambda_i} d\rho d \Omega_5 
\end{align}

The Jacobian determinant must always be included as a factor when computing norms and observables (e.g. probabilities, energies, and reaction rates) in hyperspherical coordinates.  

\subsection{Kinetic energy in Delves hyperspherical coordinates} 

The kinetic energy in Jacobi coordinates was shown to be
\begin{align}
T_i=  -\frac{\hbar^2}{2m} \left(\nabla^2_{\vec{\eta}_i} + \nabla^2_{\vec{\lambda}_i} \right )
\end{align}

Defining the total Laplacian in Jacobi coordinate $\nabla_{6D}^2=\nabla^2_{\vec{\eta}_i} + \nabla^2_{\vec{\lambda}_i}$. This 6D Laplacian can be written in hyperspherical coordinates by making use of the hyperspherical coordinate metric tensor $g = \text{diag}\left(1, \rho^2, \rho^2 \sin^2 \theta_i, \rho^2 \cos^2 \theta_i, \rho^2 \sin^2 \theta_i \sin^2 \nu_{\eta_i}, \rho^2 \cos^2 \theta_i \sin^2 \nu_{\lambda_i}\right)$. This metric is obtained from $ds^2 = d\vec{\eta}_i \cdot d\vec{\eta}_i + d\vec{\lambda}_i \cdot d\vec{\lambda}_i$, through the Delves coordinates transformation in Eq. \ref{eq:delves}. Using this metric, it can be shown that \cite{ave2018, suz1998}

\begin{align}
\nabla_{6D}^2 = \frac{\partial^2}{\partial \rho^2} + \frac{5}{\rho}\frac{\partial}{\partial \rho} + \frac{1}{\rho^2}\left[ \frac{\partial^2}{\partial \theta^2_i} + 4 \cot \theta_i \frac{\partial}{\partial \theta_i}  -\frac{\mathbf{L}^2_{\eta_i}}{\sin^2\theta_i} - \frac{\mathbf{L}^2_{\lambda_i}}{\cos^2\theta_i}\right]
\end{align}

where $\mathbf{L}_{\eta_i}$ and $\mathbf{L}_{\lambda_i}$ are the orbital angular momentum operators associated with the spherical symmetry from $\eta_i$ and $\lambda_i$, respectively. The eigenfunctions of these operators are spherical harmonics. Their eigenvalues are given by

\begin{align}
\mathbf{L}_{\eta_i}^2 Y^{\ell_{\eta_i} m_{\eta_i}}(\nu_{\eta_i},  \omega_{\eta_i}) &= \ell_{\eta_i} (\ell_{\eta_i} + 1) Y^{\ell_{\eta_i} m_{\eta_i}}(\nu_{\eta_i},  \omega_{\eta_i}) \\
\mathbf{L}_{\lambda_i}^2 Y^{\ell_{\lambda_i} m_{\lambda_i}}(\nu_{\lambda_i}, \omega_{\lambda_i}) &= \ell_{\lambda_i} (\ell_{\lambda_i} + 1) Y^{\ell_{\lambda_i} m_{\lambda_i}}(\nu_{\lambda_i}, \omega_{\lambda_i})
\end{align}

$Y^{\ell_{\eta_i} m_{\eta_i}}(\nu_{\eta_i},  \omega_{\eta_i})$ and $Y^{\ell_{\lambda_i} m_{\lambda_i}}(\nu_{\lambda_i}, \omega_{\lambda_i})$ are spherical harmonics. 

\begin{align}
Y^{\ell_{\eta_i} m_{\eta_i}}(\nu_{\eta_i}, \omega_{\eta_i}) = \sqrt{\frac{(2\ell_{\eta_i}+1)}{4\pi}\frac{(\ell_{\eta_i}-|m_{\eta_i}|)!}{(\ell_{\eta_i}+|m_{\eta_i}|)!}} \cdot P^{|m_{\eta_i}|}_{\ell_{\eta_i}}(\cos\nu_{\eta_i}) \cdot e^{im_{\eta_i}\omega_{\eta_i}} \\
Y^{\ell_{\lambda_i} m_{\lambda_i}}(\nu_{\lambda_i}, \omega_{\lambda_i}) = \sqrt{\frac{(2\ell_{\lambda_i}+1)}{4\pi}\frac{(\ell_{\lambda_i}-|m_{\lambda_i}|)!}{(\ell_{\lambda_i}+|m_{\lambda_i}|)!}} \cdot P^{|m_{\lambda_i}|}_{\ell_{\lambda_i}}(\cos\nu_{\lambda_i}) \cdot e^{im_{\lambda_i}\omega_{\lambda_i}}
\end{align}

where  $P^{|m_{\eta_i}|}_{\ell_{\eta_i}}(\cos\nu_{\eta_i})$ and $P^{|m_{\lambda_i}|}_{\ell_{\lambda_i}}(\cos\nu_{\lambda_i})$ are the associated Legendre polynomials. Since the Condon-Shortley phase factors of $(-1)^{m_{\eta_i}} $ and $(-1)^{m_{\lambda_i}} $ are not included in each spherical harmonic, they must be included in Rodrigues formula for the associated Legendre polynomials \cite{arf2012}. Different conventions exist for spherical harmonics; ensure consistency between the phase convention used and the Legendre polynomial implementation. $m_{\eta_i}$ and $m_{\lambda_i}$ are magnetic quantum numbers satisfying the conditions $-\ell_{\eta_i} \leq m_{\eta_i} \leq \ell_{\eta_i}$ and $-\ell_{\lambda_i} \leq m_{\lambda_i} \leq \ell_{\lambda_i}$, respectively. Therefore, the kinetic energy operator is

\begin{align}\label{eq:kinetic_hyperspherical}
T_i = -\frac{\hbar^2}{2m}\left (\frac{\partial^2}{\partial \rho^2} + \frac{5}{\rho}\frac{\partial}{\partial \rho} - \frac{1}{\rho^2}\mathbf{\Lambda}^2(\Omega^i_5)\right)
\end{align}

where 

\begin{align}
\mathbf{\Lambda}^2(\Omega^i_5) = -\frac{\partial^2}{\partial \theta^2_i} - 4 \cot \theta_i \frac{\partial}{\partial \theta_i}  +\frac{\mathbf{L}^2_{\eta_i}}{\sin^2\theta_i} + \frac{\mathbf{L}^2_{\lambda_i}}{\cos^2\theta_i}  
\end{align}

It can be clearly observed that the kinetic energy operator has been separated into hyperradial and angular parts. This is one of the main reasons for using hyperspherical coordinates. The operator $\mathbf{\Lambda}(\Omega^i_5)$ is called the grand angular momentum operator, or sometimes the hyperangular momentum. Its square, $\mathbf{\Lambda}^2(\Omega^i_5)$, describes the angular part of the kinetic energy. The symmetry associated with this operator is the rotational symmetry on the 5-sphere. $\mathbf{\Lambda}^2(\Omega^i_5)$ is the Laplace-Beltrami operator on the 5-sphere $S^5$.

\subsection{Faddeev equations in hyperspherical coordinates}

Two-body wavefunctions in Jacobi coordinates $\psi_i(\vec{\eta}_i, \vec{\lambda}_i)$ are written as $\psi_i(\rho, \Omega^i_5)$ in hyperspherical coordinates. Using the kinetic energy operator in hyperspherical coordinates (Eq. \ref{eq:kinetic_hyperspherical}), Faddeev equations in Jacobi coordinates (Eq. \ref{eq:faddeev1}) now transform to

\begin{align}
\label{eq:faddev_hyperspherical}
\left[- \frac{\hbar^2}{2m}\left(\frac{\partial^2}{\partial \rho^2} + \frac{5}{\rho}\frac{\partial}{\partial \rho} - \frac{1}{\rho^2}\mathbf{\Lambda}^2(\Omega^i_5)\right) + V_i(\rho, \Omega^i_5) - E\right]\psi_i(\rho, \Omega^i_5) = -V_i(\rho, \Omega^i_5)[\psi_j(\rho, \Omega^j_5) + \psi_k(\rho, \Omega^k_5)]
\end{align}

The potential $V_i(\rho, \Omega^i_5)$ simplifies to $V_i(\rho, \theta_i)$ in hyperspherical coordinates. This dependence follows directly from the Faddeev decomposition, where in each partition there is a spectator and only two interacting particles, whose interaction depends on their relative separation. For spectator $i$, the separation between $j$ and $k$ depends on $\eta_i = \rho \sin \theta_i$. The potential now has a more complex functional form since it depends on two variables, unlike the case in Jacobi coordinates where it depends only on one variable $V_i(\eta_i)$. The quantum numbers that label the states of the system are now built into the two-body wavefunctions. Let $\alpha_i$ be the set of these quantum numbers with their proper coupling order to give the state $J=L+S$, where $L$ and $S$ are the orbital angular momentum and the spin of the three-body system, respectively. The wavefunctions are now written as $\psi^{J\alpha_i}_i(\rho, \Omega^i_5)$. Since the kinetic energy is already separated into hyperradial and angular parts, this \textit{suggests} that the two-body wavefunctions $\psi^{J\alpha_i}_i(\rho, \Omega^i_5)$ should be separated in a similar manner. The eigenfunctions of the square of the grand angular momentum operator, $\mathbf{\Lambda}^2(\Omega^i_5)$, are called hyperspherical harmonics $\mathcal{Y}^{\ell_{\eta_i} \ell_{\lambda_i}}_{K_i} (\Omega^i_5)$. The action of $\mathbf{\Lambda}^2(\Omega^i_5)$ on this angular wavefunction is as follows \cite{ave2018}:  

\begin{align}
\mathbf{\Lambda}^2(\Omega^i_5) \mathcal{Y}^{\ell_{\eta_i} \ell_{\lambda_i}}_{K_i} (\Omega^i_5) = K_i(K_i+4) \mathcal{Y}^{\ell_{\eta_i} \ell_{\lambda_i}}_{K_i}(\Omega^i_5)
\end{align}

where $K_i(K_i+4)$ are the eigenvalues, with $K_i$ being the hyperangular momentum quantum number. Hyperspherical harmonics are constructed by coupling spherical harmonics \(Y^{\ell_{\eta_i} m_{\eta_i}}\) and \(Y^{\ell_{\lambda_i} m_{\lambda_i}}\) through the hyperangle \(\theta_i\) \cite{tho2009a}: 

\begin{align}
\label{eq:hyperangular}
\mathcal{Y}^{\ell_{\eta_i} \ell_{\lambda_i}}_{K_i}(\Omega^i_5)= \phi_{K_i}^{\ell_{\eta_i}, \ell_{\lambda_i}}(\theta_i) \left[ Y^{\ell_{\eta_i} m_{\eta_i}} \otimes Y^{\ell_{\lambda_i} m_{\lambda_i}} \right]_{LM},
\end{align}

where $\phi_{K_i}^{\ell_{\eta_i}, \ell_{\lambda_i}}(\theta_i)$ is a hyperangular polynomial given by

\begin{align} \label{eq:hyper_polynomial}
\phi_{K_i}^{\ell_{\eta_i}, \ell_{\lambda_i}}(\theta_i) = N_{K_i}^{\ell_{\eta_i}, \ell_{\lambda_i}} \, (\sin \theta_i)^{\ell_{\eta_i}} (\cos \theta_i)^{\ell_{\lambda_i}} P_{n_i}^{(\ell_{\lambda_i} + \frac{1}{2}, \ell_{\eta_i} + \frac{1}{2})}(\cos 2\theta_i),
\end{align}

where $P^{(a,b)}_{n_i}(x)$ is the Jacobi polynomial. The normalisation constant $N_{K}^{\ell_{\eta_i}, \ell_{\lambda_i}} $ is given by

\begin{align}
N_{K_i}^{\ell_{\eta_i}, \ell_{\lambda_i}} = \sqrt{\frac{2(2n_i+\ell_{\eta_i}+\ell_{\lambda_i}+2) \, n_i! \, \Gamma(n_i+\ell_{\eta_i}+\ell_{\lambda_i}+2)}{\Gamma(n_i+\ell_{\eta_i}+\frac{3}{2}) \, \Gamma(n_i+\ell_{\lambda_i}+\frac{3}{2})}}
\end{align}
The quantum number $K_i$, $ \ell_{\eta_i}$   and $\ell_{\lambda_i}$  have the following restrictions: 

\begin{enumerate}
\item $K_i$ is a nonnegative integer i.e. $K_i=0,1,2, ...$. Its values are further restricted by the next two constraints.
\item $K_i \geq \ell_{\eta_i} +\ell_{\lambda_i}$. The minimum value of $K_i$ is determined by $\ell_{\eta_i}$ and $\ell_{\lambda_i}$.
\item $K_i - (\ell_{\eta_i} +\ell_{\lambda_i})$ must be even. This implies $K_i - (\ell_{\eta_i} +\ell_{\lambda_i})=2n_i$, where $n_i=0,1,2,3, ...$
\end{enumerate}
The tensor product $\left[ Y^{\ell_{\eta_i} m_{\eta_i}} \otimes Y^{\ell_{\lambda_i} m_{\lambda_i}} \right]_{LM}$ represents the angular momentum coupling of two spherical harmonics to form a combined state with total angular momentum $L$ and magnetic quantum number $M$. In other words, one is given the orbital angular momentum $\ell_{\eta_i}$ with its projection $m_{\eta_i}$ and the orbital angular momentum $\ell_{\lambda_i}$ with its projection $m_{\lambda_i}$ to construct the state with total angular momentum $L$ and magnetic quantum number $M$. Explicitly, this coupled state can be written as summations of products of spherical harmonics over the magnetic quantum numbers:

\begin{align}\label{eq:clebsch}
\left[ Y^{\ell_{\eta_i} m_{\eta_i}} \otimes Y^{\ell_{\lambda_i} m_{\lambda_i}} \right]_{LM} = \sum_{m_{\eta_i},m_{\lambda_i}} C^{LM}_{\ell_{\eta_i}m_{\eta_i},\ell_{\lambda_i}m_{\lambda_i}} \, Y^{\ell_{\eta_i} m_{\eta_i}}(\nu_{\eta_i}, \omega_{\eta_i}) \, Y^{\ell_{\lambda_i} m_{\lambda_i}}(\nu_{\lambda_i}, \omega_{\lambda_i})
\end{align}

where $C^{LM}_{\ell_{\eta_i}m_{\eta_i},\ell_{\lambda_i}m_{\lambda_i}}=\langle \ell_{\eta_i} m_{\eta_i}, \ell_{\lambda_i} m_{\lambda_i} | LM \rangle$ are the Clebsch-Gordan coefficients. The magnetic quantum number $M$ ranges from $-L$ to $L$ in integer steps. Two important restrictions (selection rules) govern this coupling \cite{mes1962}

\begin{enumerate}
\item The total angular momentum quantum number $L$ must satisfy the triangular inequality: $|\ell_{\eta_i} - \ell_{\lambda_i}| \leq L \leq \ell_{\eta_i} + \ell_{\lambda_i}$
\item Angular momentum conservation requires $m_{\eta_i} + m_{\lambda_i} = M$
\end{enumerate}

If these two conditions are not met the Clebsch-Gordan coefficients are equal to zero. These constraints guide the selection of valid quantum numbers when constructing wavefunctions. It is usually preferable to replace the Clebsch-Gordan coefficients with Wigner 3j symbols, so that the symmetry properties of the Wigner 3j symbols may be exploited.  In Fano-Racah notation, the Clebsch-Gordan coefficients are replaced by the coefficients $\langle \ell_{\eta_i} m_{\eta_i}, \ell_{\lambda_i} m_{\lambda_i} | LM \rangle=C({\ell_{\eta_i} \ell_{\lambda_i} L|m_{\eta_i}m_{\lambda_i}} LM)$, where

\begin{align}
C({\ell_{\eta_i} \ell_{\lambda_i} L|m_{\eta_i}m_{\lambda_i}} LM) = (-1)^{\ell_{\eta_i}-\ell_{\lambda_i}+M} \sqrt{2L + 1} \begin{pmatrix} \ell_{\eta_i} & \ell_{\lambda_i} & L \\ m_{\eta_i} & m_{\lambda_i} & -M \end{pmatrix}
\end{align}

where
\begin{align}
\begin{pmatrix} \ell_{\eta_i} & \ell_{\lambda_i} & L \\ m_{\eta_i} & m_{\lambda_i} & -M \end{pmatrix}
\end{align}

is the Wigner 3j symbol. The conservation of angular momentum requires $m_{\eta_i} + m_{\lambda_i} = M$ for the Clebsch-Gordan coefficient, and correspondingly $m_{\eta_i} + m_{\lambda_i} - M = 0$ for the 3j symbols. Wigner 3j symbols are computed using the Racah formula in terms of factorials and phase factors \cite{mes1962}. There exist tables of calculated Wigner 3j symbols; see, for example, \cite{rot1959}. Most programming languages have libraries for computing spherical harmonics, Clebsch-Gordan coefficients and Wigner 3j symbols.

\subsection{Ground state of the triton}

The triton is a three-body system that consists of one proton and two neutrons. The ground state of the triton is mainly $L=0$ (S-wave), with a small admixture of $L=2$ (D-wave). This mixing of S and D states is caused by tensor forces. The ground state of the triton is about 90\% S-waves and about 10\% D-waves. For each of these states, a coupling scheme for $L$ is illustrated and hyperspherical harmonics constructed. Let $\vec{\eta_i}$ be the relative vector between the two neutrons, and $\vec{\lambda_i}$ the relative vector from the proton to the centre-of-mass of the two neutrons. 

\begin{enumerate}[(1)]
\item $L=0$ state \\

$L$ is a vector sum of $\ell_{\eta_i}$ and $\ell_{\lambda_i}$. Therefore, the only possible orbital angular momenta for $L=0$ are $\ell_{\eta_i}=0$ and $\ell_{\lambda_i}=0$ i.e. $L=\ell_{\eta_i} + \ell_{\lambda_i}=0$.

        \begin{enumerate}
            \item For $L=0$, only $M=0$ is allowed
          \item For $\ell_{\eta_i}=0$, the only allowed magnetic quantum number is $m_{\eta_i}=0$
          \item For $\ell_{\lambda_i}=0$, the only allowed magnetic quantum number is also $m_{\lambda_i}=0$
          \item $M=m_{\eta_i}+m_{\lambda_i}=0$
          \item $Y^{\ell_{\eta_i} m_{\eta_i}}=Y^{00}$   
          \item $Y^{\ell_{\lambda_i} m_{\lambda_i}}=Y^{00}$ 
          \item Making use of Eq. \ref{eq:clebsch},
           \begin{align}
\left[ Y^{0 m_{\eta_i}} \otimes Y^{0 m_{\lambda_i}} \right]_{00} = C(000|000) \, Y^{00}(\nu_{\eta_i}, \omega_{\eta_i}) \, Y^{00}(\nu_{\lambda_i}, \omega_{\lambda_i})
\end{align}

where 

\begin{align}
C(000|000)= (-1)^{0-0+0} \sqrt{2(0) + 1} 
\begin{pmatrix} 0 & 0 & 0 \\
0 & 0 & 0 \end{pmatrix} 
= \begin{pmatrix} 0 & 0 & 0 \\
0 & 0 & 0 \end{pmatrix}
=1
\end{align}

\item The hyperangular polynomials (From Eq. \ref{eq:hyper_polynomial}) are 

\begin{align} 
\phi_{K_i}^{0,0}(\theta_i) = N_{K_i}^{0, 0} \, (\sin \theta_i)^{0} (\cos \theta_i)^{0} P_{n_i}^{( \frac{1}{2}, \frac{1}{2})}(\cos 2\theta_i) =
N_{K_i}^{0, 0} P_{n_i}^{( \frac{1}{2}, \frac{1}{2})}(\cos 2\theta_i)
\end{align}

         \item Putting everything together, the hyperspherical harmonic for the state $L=0, M=0$ is
         
        \begin{align}
\mathcal{Y}^{0 0}_{K_i}(\Omega^i_5)&= \phi_{K_i}^{0, 0}(\theta_i) \, Y^{00}(\nu_{\eta_i}, \omega_{\eta_i}) \, Y^{00}(\nu_{\lambda_i}, \omega_{\lambda_i}) \\
       &= \phi_{K_i}^{0, 0}(\theta_i). \frac{1}{\sqrt{4\pi}} .   \frac{1}{\sqrt{4\pi}} \\
       &= \frac{1}{4 \pi} \phi_{K_i}^{0, 0}(\theta_i)
\end{align}
                   \item $K_i = \ell_{\eta_i} +\ell_{\lambda_i} +2n_i=0 + 0+2n_i$. Therefore, $K_i=0,2,4, 6 ...$ are the allowed values.
         \end{enumerate}
\item $L=2$ state\\

Possible sets of orbital angular momenta for $L=2$ are $(\ell_{\eta_i}, \ell_{\lambda_i})=(2,0)$,  $(\ell_{\eta_i}, \ell_{\lambda_i})=(1,1)$ and $(\ell_{\eta_i}, \ell_{\lambda_i})=(0,2)$. The choice $(\ell_{\eta_i}, \ell_{\lambda_i})=(2,0)$ is used.
       \begin{enumerate}
       \item  For $L=2$, $M=-2,-1,0,1,2$ are allowed. A specific choice of $M$ is made, usually based on the physics. In this example $M=0$ is used.
        \item For $\ell_{\eta_i}=2$, the allowed magnetic quantum numbers are $m_{\eta_i}=-2,-1,0,1,2$
          \item For $\ell_{\lambda_i}=0$, the only allowed magnetic quantum number is $m_{\lambda_i}=0$
          \item Since $M=0$, $M=m_{\eta_i}+m_{\lambda_i}=m_{\eta_i}+0=0$. This implies that only $m_{\eta_i}=0$ is allowed, from all the possible values. 
          \item $Y^{\ell_{\eta_i} m_{\eta_i}}=Y^{20}$.   
          \item $Y^{\ell_{\lambda_i} m_{\lambda_i}}=Y^{00}$ 
          \item \begin{align}
\left[ Y^{20} \otimes Y^{00} \right]_{20} &= \sum_{m_{\eta_i}=-2}^{2} \sum_{m_{\lambda_i}=0} C(2 0 2|m_{\eta_i}m_{\lambda_i} M) \, Y^{\ell_{\eta_i} m_{\eta_i}}(\nu_{\eta_i}, \omega_{\eta_i}) \, Y^{\ell_{\lambda_i} m_{\lambda_i}}(\nu_{\lambda_i}, \omega_{\lambda_i}) \\
&=  C(2 02|00 0) \, Y^{20}(\nu_{\eta_i}, \omega_{\eta_i}) \, Y^{0 0}(\nu_{\lambda_i}, \omega_{\lambda_i}) 
\end{align}

\begin{align}
C(202|000)&= (-1)^{2-0+0} \sqrt{2(2) + 1} 
\begin{pmatrix} 2 & 0 & 2 \\
0 & 0 & 0 \end{pmatrix} \\
&= \sqrt{5} \begin{pmatrix} 2 & 0 & 2 \\
0 & 0 & 0 \end{pmatrix}
=\sqrt{5} \left( \frac{1}{\sqrt{5} } \right)= 1
\end{align}
\item The hyperangular polynomials are 

\begin{align} 
\phi_{K_i}^{2,0}(\theta_i) = N_{K_i}^{2, 0} \, (\sin \theta_i)^{2} (\cos \theta_i)^{0} P_{n_i}^{( \frac{5}{2}, \frac{1}{2})}(\cos 2\theta_i) = N_{K_i}^{2, 0} \, (\sin \theta_i)^{2} P_{n_i}^{( \frac{5}{2}, \frac{1}{2})}(\cos 2\theta_i)
\end{align}

\item  Hyperspherical harmonics for $L=2, M=0$ are therefore
         
        \begin{align}
\mathcal{Y}^{2 0}_{K_i}(\Omega^i_5) &= \phi_{K_i}^{2, 0}(\theta_i) Y^{20}(\nu_{\eta_i}, \omega_{\eta_i}) \, Y^{0 0}(\nu_{\lambda_i}, \omega_{\lambda_i}) \\
   &= \phi_{K_i}^{2, 0}. \sqrt{\frac{5}{16\pi}} \left( 3\cos^2 \nu_{\eta_i} - 1 \right). \frac{1}{\sqrt{4\pi}} \\
   &= \phi_{K_i}^{2, 0}. \dfrac{\sqrt{5}}{8\pi} \left( 3\cos^2 \nu_{\eta_i} - 1 \right)
\end{align}
\item $K_i = \ell_{\eta_i} +\ell_{\lambda_i} +2n_i=2 + 0+2n_i$. Therefore, $K_i=2,4,6, ...$.
       \end{enumerate}
\end{enumerate}

\section{Coupled-channel hyperradial equations}

\subsection{Eigenfunction expansion}
Hyperspherical harmonics form an angular basis for functions on the five-dimensional hypersphere because they are orthonormal and complete. As a way to separate the two-body wavefunction into hyperradial and angular parts, it is therefore a good strategy to expand the wavefunction on this angular basis. The coefficients on this basis are hyperradial wavefunctions.

\begin{align}
\label{eq:expansion1}
\psi^{J\alpha_i}_i(\rho, \Omega^i_5) =  \sum_{K_i} \frac{\chi_{K_i}^{i,J \alpha_i}(\rho)}{\rho^{5/2}} \mathcal{Y}^{\ell_{\eta_i} \ell_{\lambda_i}}_{K_i}(\Omega^i_5)
\end{align}

The factor $1/\rho^{5/2} $ is included for reasons that will become clear as this discussion progresses. This expansion in Eq. \ref{eq:expansion1} is truncated at a suitable value $K_{max}$, after convergence is achieved. The expansion is substituted into the Faddeev equations in hyperspherical coordinates (Eq. \ref{eq:faddev_hyperspherical}).

\begin{align}
\left[- \frac{\hbar^2}{2m}\left(\frac{\partial^2}{\partial \rho^2} + \frac{5}{\rho}\frac{\partial}{\partial \rho} -\frac{1}{\rho^2}\mathbf{\Lambda}^2(\Omega^i_5)\right) + V_i(\rho, \Omega^i_5) - E\right] \sum_{K_i} \frac{\chi_{K_i}^{i,J \alpha_i}(\rho)}{\rho^{5/2}} \mathcal{Y}^{\ell_{\eta_i} \ell_{\lambda_i}}_{K_i}(\Omega^i_5)   \notag \\
= -V_i(\rho, \Omega^i_5)\sum_{K_j} \frac{\chi_{K_j}^{j,J \alpha_j}(\rho)}{\rho^{5/2}} \mathcal{Y}^{\ell_{\eta_j} \ell_{\lambda_j}}_{K_j}(\Omega^j_5) 
-V_i(\rho, \Omega^i_5) \psi_k(\rho, \Omega^k_5) \sum_{K_k} \frac{\chi_{K_k}^{k,J \alpha_k}(\rho)}{\rho^{5/2}} \mathcal{Y}^{\ell_{\eta_k} \ell_{\lambda_k}}_{K_k}(\Omega^k_5)
\end{align}

This equation looks very complex. The action of the operators on the wave function shall be carried out one term at a time, for the sake of clarity.

\begin{enumerate}[(1)]
\item Hyperradial differentiation term

First derivative:

\begin{align}
\frac{d}{d\rho}\left(\frac{\chi_{K_i}^{i,J \alpha_i}(\rho)}{\rho^{5/2}}\right) &= \frac{1}{\rho^{5/2}}\frac{d\chi_{K_i}^{i,J \alpha_i}(\rho)}{d\rho} - \frac{5}{2}\frac{\chi_{K_i}^{i,J \alpha_i}(\rho)}{\rho^{7/2}} \notag \\
\frac{5}{\rho}\frac{d}{d\rho}\left(\frac{\chi_{K_i}^{i,J \alpha_i}(\rho)}{\rho^{5/2}}\right) &= \frac{5}{\rho}\left[\frac{1}{\rho^{5/2}}\frac{d\chi_{K_i}^{i,J \alpha_i}(\rho)}{d\rho} - \frac{5}{2}\frac{\chi_{K_i}^{i,J \alpha_i}(\rho)}{\rho^{7/2}}\right] \notag \\
&= \frac{5}{\rho^{7/2}}\frac{d\chi_{K_i}^{i,J \alpha_i}(\rho)}{d\rho} - \frac{25}{2}\frac{\chi_{K_i}^{i,J \alpha_i}(\rho)}{\rho^{9/2}}
\end{align}

Second derivative:

\begin{align}
\frac{d^2}{d\rho^2}\left(\frac{\chi_{K_i}^{i,J \alpha_i}(\rho)}{\rho^{5/2}}\right) &= \frac{d}{d\rho}\left(\frac{1}{\rho^{5/2}}\frac{d\chi_{K_i}^{i,J \alpha_i}(\rho)}{d\rho} - \frac{5}{2}\frac{\chi_{K_i}^{i,J \alpha_i}(\rho)}{\rho^{7/2}}\right) \notag \\
&= \frac{1}{\rho^{5/2}}\frac{d^2\chi_{K_i}^{i,J \alpha_i}(\rho)}{d\rho^2} - \frac{5}{2}\frac{1}{\rho^{7/2}}\frac{d\chi_{K_i}^{i,J \alpha_i}(\rho)}{d\rho} - \frac{5}{2}\frac{d}{d\rho}\left(\frac{\chi_{K_i}^{i,J \alpha_i}(\rho)}{\rho^{7/2}}\right)  \notag\\
&= \frac{1}{\rho^{5/2}}\frac{d^2\chi_{K_i}^{i,J \alpha_i}(\rho)}{d\rho^2} - \frac{5}{2}\frac{1}{\rho^{7/2}}\frac{d\chi_{K_i}^{i,J \alpha_i}(\rho)}{d\rho} - \frac{5}{2}\left[\frac{1}{\rho^{7/2}}\frac{d\chi_{K_i}^{i,J \alpha_i}(\rho)}{d\rho} - \frac{7}{2}\frac{\chi_{K_i}^{i,J \alpha_i}(\rho)}{\rho^{9/2}}\right] \notag \\
&= \frac{1}{\rho^{5/2}}\frac{d^2\chi_{K_i}^{i,J \alpha_i}(\rho)}{d\rho^2} - \frac{5}{\rho^{7/2}}\frac{d\chi_{K_i}^{i,J \alpha_i}(\rho)}{d\rho} + \frac{35}{4}\frac{\chi_{K_i}^{i,J \alpha_i}(\rho)}{\rho^{9/2}}
\end{align}

Combining the first and second derivatives:

\begin{align}
\left(\frac{d^2}{d\rho^2} + \frac{5}{\rho}\frac{d}{d\rho}\right)\left(\frac{\chi_{K_i}^{i,J \alpha_i}(\rho)}{\rho^{5/2}}\right) =
& \frac{1}{\rho^{5/2}}\frac{d^2\chi_{K_i}^{i,J \alpha_i}(\rho)}{d\rho^2} - \frac{5}{\rho^{7/2}}\frac{d\chi_{K_i}^{i,J \alpha_i}(\rho)}{d\rho} \notag \\
&+ \frac{35}{4}\frac{\chi_{K_i}^{i,J \alpha_i}(\rho)}{\rho^{9/2}} + \frac{5}{\rho^{7/2}}\frac{d\chi_{K_i}^{i,J \alpha_i}(\rho)}{d\rho} - \frac{25}{2}\frac{\chi_{K_i}^{i,J \alpha_i}(\rho)}{\rho^{9/2}} \notag \\
&= \frac{1}{\rho^{5/2}}\frac{d^2\chi_{K_i}^{i,J \alpha_i}(\rho)}{d\rho^2} + \left(\frac{35}{4} - \frac{25}{2}\right)\frac{\chi_{K_i}^{i,J \alpha_i}(\rho)}{\rho^{9/2}} \notag \\
&= \frac{1}{\rho^{5/2}}\frac{d^2\chi_{K_i}^{i,J \alpha_i}(\rho)}{d\rho^2} + \left(\frac{35}{4} - \frac{50}{4}\right)\frac{\chi_{K_i}^{i,J \alpha_i}(\rho)}{\rho^{9/2}} \notag \\
&= \frac{1}{\rho^{5/2}}\frac{d^2\chi_{K_i}^{i,J \alpha_i}(\rho)}{d\rho^2} - \frac{15}{4}\frac{\chi_{K_i}^{i,J \alpha_i}(\rho)}{\rho^{9/2}} 
\end{align}

The hyperradial differential term may now completely be written as

\begin{align}
\left(\frac{\partial ^2}{\partial \rho^2} + \frac{5}{\rho}\frac{\partial}{\partial \rho}\right) \left( \sum_{K_i}\frac{\chi_{K_i}^{i,J \alpha_i}(\rho)}{\rho^{5/2}} \mathcal{Y}^{\ell_{\eta_i} \ell_{\lambda_i}}_{K_i}(\Omega^i_5) \right) = \notag \\
 \sum_{K_i}\left ( \frac{d^2\chi_{K_i}^{i,J \alpha_i}(\rho)}{d\rho^2} - \frac{15}{4}\frac{\chi_{K_i}^{i,J \alpha_i}(\rho)}{\rho^{2}}  \right ) \frac{\mathcal{Y} ^{\ell_{\eta_i} \ell_{\lambda_i}}_{K_i}(\Omega^i_5)}{\rho^{5/2}} 
\end{align}

It can be observed that the first derivative that was present in the Faddeev equations in hyperspherical coordinates (Eq. \ref{eq:faddev_hyperspherical}) has disappeared after this procedure of differentiation. This was made possible because of the factor $1/\rho^{5/2}$ included in the expansion. First derivatives are non-hermitian, giving rise to matrices that are non-symmetric after discretisation: this is known to cause instabilities in many numerical methods. Therefore, in the absence of the first derivative, the final equation shall look Schr\"{o}dinger-like, making it possible to use well-established computational techniques for the Schr\"{o}dinger equation. 

\item Grand angular momentum operator term

\begin{align}
-\frac{1}{\rho^2}\mathbf{\Lambda}^2(\Omega^i_5) \left (\sum_{K_i} \frac{\chi_{K_i}^{i,J \alpha_i}(\rho)}{\rho^{5/2}}\mathcal{Y}^{\ell_{\eta_i} \ell_{\lambda_i}}_{K_i}(\Omega^i_5)\right) 
&= -\sum_{K_i} \frac{\chi_{K_i}^{i,J \alpha_i}(\rho)}{\rho^{5/2}} \cdot \frac{K_i(K_i+4)}{\rho^2} \mathcal{Y}^{\ell_{\eta_i} \ell_{\lambda_i}}_{K_i}(\Omega^i_5) \notag \\
&= -\sum_{K_i}\frac{K_i(K_i+4)}{\rho^{9/2}}\chi_{K_i}^{i,J \alpha_i}(\rho) \mathcal{Y}^{\ell_{\eta_i} \ell_{\lambda_i}}_{K_i}(\Omega^i_5)
\end{align}
\end{enumerate}

Combining the hyperradial differential term and the grand angular momentum term, the left hand side of the Faddeev equation becomes

\begin{align}
& \sum_{K_i}\left[ -\frac{\hbar^2}{2m}\left(\frac{d^2\chi_{K_i}^{i,J \alpha_i}(\rho)}{d\rho^2} - \frac{15}{4\rho^{2}}\chi_{K_i}^{i,J \alpha_i}(\rho) - \frac{K_i(K_i+4)}{\rho^{2}}\chi_{K_i}^{i,J \alpha_i}(\rho) \right) +  
  (V_i(\rho, \Omega^i_5) - E)\chi_{K_i}^{i,J \alpha_i}(\rho) \right] \frac{\mathcal{Y}^{\ell_{\eta_i} \ell_{\lambda_i}}_{K_i}(\Omega^i_5)}{\rho^{5/2}} \notag \\
&=\sum_{K_i}\left[-\frac{\hbar^2}{2m}  \left(    \frac{d^2\chi_{K_i}^{i,J \alpha_i}(\rho)}{d\rho^2}  - \frac{K_i(K_i+4) + 15/4}{\rho^{2}}\chi_{K_i}^{i,J \alpha_i}(\rho)   \right) + 
  (V_i(\rho, \Omega^i_5) - E)\chi_{K_i}^{i,J \alpha_i}(\rho) \right] \frac{\mathcal{Y}^{\ell_{\eta_i} \ell_{\lambda_i}}_{K_i}(\Omega^i_5)}{\rho^{5/2}}
\end{align}

Define 

\begin{align}
L_{K_i}= \frac{\hbar^2 (K_i(K_i+4) + 15/4)}{2m\rho^{2}}
\end{align}
This is called the centrifugal barrier. The left hand side of Faddeev equations can now be written as 

\begin{align}
 \sum_{K_i}\left[-\frac{\hbar^2}{2m}\frac{d^2\chi_{K_i}^{i,J \alpha_i}(\rho)}{d\rho^2}  +L_{K_i} 
  +(V_i(\rho, \Omega^i_5) - E)\chi_{K_i}^{i,J \alpha_i}(\rho) \right] \frac{\mathcal{Y}^{\ell_{\eta_i} \ell_{\lambda_i}}_{K_i}(\Omega^i_5)}{\rho^{5/2}}
\end{align}

Including the right-hand side, Faddeev equations in hyperspherical coordinates are given by

\begin{align}
 \sum_{K_i}\left[-\frac{\hbar^2}{2m}\frac{d^2\chi_{K_i}^{i,J \alpha_i}(\rho)}{d\rho^2}  +L_{K_i} 
  +(V_i(\rho, \Omega^i_5) - E)\chi_{K_i}^{i,J \alpha_i}(\rho) \right] \frac{\mathcal{Y}^{\ell_{\eta_i} \ell_{\lambda_i}}_{K_i}(\Omega^i_5)}{\rho^{5/2}} = \notag \\
   -V_i(\rho, \Omega^i_5)\sum_{K_j} \frac{\chi_{K_j}^{j,J \alpha_j}(\rho)}{\rho^{5/2}} Y^{\ell_{\eta_j} \ell_{\lambda_j}}_{K_j}(\Omega^j_5) 
-V_i(\rho, \Omega^i_5) \sum_{K_k} \frac{\chi_{K_k}^{k,J \alpha_k}(\rho)}{\rho^{5/2}} \mathcal{Y}^{\ell_{\eta_k} \ell_{\lambda_k}}_{K_k}(\Omega^k_5)
\end{align}

The factor $\frac{1}{\rho^{5/2}} $ is canceled from both side, giving rise to

\begin{align}
 \sum_{K_i}\left[-\frac{\hbar^2}{2m}\frac{d^2\chi_{K_i}^{i,J \alpha_i}(\rho)}{d\rho^2}  +L_{K_i} 
  +(V_i(\rho, \Omega^i_5) - E)\chi_{K_i}^{i,J \alpha_i}(\rho) \right] \mathcal{Y}^{\ell_{\eta_i} \ell_{\lambda_i}}_{K_i}(\Omega^i_5) = \notag \\
   -V_i(\rho, \Omega^i_5)\sum_{K_j} \chi_{K_j}^{j,J \alpha_j}(\rho) \mathcal{Y}^{\ell_{\eta_j} \ell_{\lambda_j}}_{K_j}(\Omega^j_5) 
-V_i(\rho, \Omega^i_5) \sum_{K_k} \chi_{K_k}^{k,J \alpha_k}(\rho) \mathcal{Y}^{\ell_{\eta_k} \ell_{\lambda_k}}_{K_k}(\Omega^k_5)
\end{align}

Keeping potential terms to one side, one arrives at

\begin{align}\label{eq:faddeev_coupled_unprojected}
 &\sum_{K_i}\left[-\frac{\hbar^2}{2m}\frac{d^2}{d\rho^2}  +L_{K_i} 
   - E \right] \chi_{K_i}^{i,J \alpha_i}(\rho)\mathcal{Y}^{\ell_{\eta_i} \ell_{\lambda_i}}_{K_i}(\Omega^i_5) = \notag \\
&   - V_i(\rho, \Omega^i_5) \sum_{K_i} \chi_{K_i}^{i,J \alpha_i}(\rho)\mathcal{Y}^{\ell_{\eta_i} \ell_{\lambda_i}}_{K_i}(\Omega^i_5) \notag \\
&   -V_i(\rho, \Omega^i_5)\sum_{K_j} \chi_{K_j}^{j,J \alpha_j}(\rho) \mathcal{Y}^{\ell_{\eta_j} \ell_{\lambda_j}}_{K_j}(\Omega^j_5) 
-V_i(\rho, \Omega^i_5) \sum_{K_k} \chi_{K_k}^{k,J \alpha_k}(\rho) \mathcal{Y}^{\ell_{\eta_k} \ell_{\lambda_k}}_{K_k}(\Omega^k_5)
\end{align}

\subsection{Projection onto hyperspherical harmonics basis}

The Faddeev equations in Eq. \ref{eq:faddeev_coupled_unprojected} are now projected onto the hyperspherical harmonics basis corresponding to the spectator. Using spectator basis element $\mathcal{Y}^{\ell_{\eta_i} \ell_{\lambda_i}}_{K_i}(\Omega^i_5)$, both side of the equation are multiplied by $\left [\mathcal{Y}^{\ell_{\eta_i} \ell_{\lambda_i}}_{K_i}(\Omega^i_5) \right ]^{\ast}$ and then integrate over the entire angular space. For clarity, the left-hand side and the right-hand side shall be handled separately. The left-hand side becomes 

\begin{align}\label{eq:faddeev_coupled2}
 \sum_{K'_i}\left[-\frac{\hbar^2}{2m}\frac{d^2}{d\rho^2}  +L_{K'_i}(\rho) 
  - E) \right] \chi_{K'_i}^{i,J \alpha_i}(\rho) \int  \left [\mathcal{Y} ^{\ell_{\eta_i} \ell_{\lambda_i}}_{K_i}(\Omega^i_5) \right ]^{\ast} \mathcal{Y}^{\ell_{\eta_i} \ell_{\lambda_i}}_{K'_i}(\Omega^i_5) d \Omega^i_5 
\end{align}

The orthonormal relations for spherical harmonics is as follows \cite{ave2018}:

\begin{align}
 \int \left [\mathcal{Y}^{\ell_{\eta_i} \ell_{\lambda_i}}_{K_i}(\Omega^i_5) \right ]^{\ast} \mathcal{Y}^{\ell_{\eta_i} \ell_{\lambda_i}}_{K'_i}(\Omega^i_5) d \Omega^i_5 = \delta_{K_i K'_i} 
\end{align}

Applying this orthonormality condition and using the sifting property of the Kronecker delta, this simplifies to the form

\begin{align}\label{eq:faddeev_coupled4}
 \sum_{K'_i}\left[-\frac{\hbar^2}{2m}\frac{d^2}{d\rho^2}  +L_{K'_i}-E 
   \right]   \delta_{K_i K'_i} \chi_{K'_i}^{i,J \alpha_i}(\rho) = \notag \\
\left[-\frac{\hbar^2}{2m}\frac{d^2}{d\rho^2}  +L_{K_i} -E
   \right]    \chi_{K_i}^{i,J \alpha_i}(\rho)
\end{align}

The right-hand side contains three potential terms. These are projected individually as follows:
\begin{enumerate}[(1)]
\item First term: $-V_i(\rho, \Omega^i_5) \sum_{K_i} \chi_{K_i}^{i,J \alpha_i}(\rho)\mathcal{Y}^{\ell_{\eta_i} \ell_{\lambda_i}}_{K_i}(\Omega^i_5)$

The projection is:
\begin{align}
 -\int [\mathcal{Y}^{\ell_{\eta_i} \ell_{\lambda_i}}_{K_i}(\Omega^i_5)]^* V_i(\rho, \Omega^i_5) \sum_{K'_i} \chi_{K'_i}^{i,J \alpha_i}(\rho)\mathcal{Y}^{\ell_{\eta_i} \ell_{\lambda_i}}_{K'_i}(\Omega^i_5) d\Omega^i_5   
\end{align}

Moving the summation and hyperradial function outside the angular integration

\begin{align}
-\sum_{K'_i} \chi_{K'_i}^{i,J \alpha_i}(\rho) \int [\mathcal{Y}^{\ell_{\eta_i} \ell_{\lambda_i}}_{K_i}(\Omega^i_5)]^* V_i(\rho, \Omega^i_5) \mathcal{Y}^{\ell_{\eta_i} \ell_{\lambda_i}}_{K'_i}(\Omega^i_5) d\Omega^i_5    
\end{align}

The diagonal coupling potential is defined by this integral

\begin{align}
V^{ii}_{K_iK'_i}(\rho) = \int [\mathcal{Y}^{\ell_{\eta_i} \ell_{\lambda_i}}_{K_i}(\Omega^i_5)]^* V_i(\rho, \Omega^i_5) \mathcal{Y}^{\ell_{\eta_i} \ell_{\lambda_i}}_{K'_i}(\Omega^i_5) d\Omega^i_5
\end{align}

Therefore, the projected term may be written as 

\begin{align}
-\sum_{K'_i} V^{ii}_{K_iK'_i}(\rho) \chi_{K'_i}^{i,J \alpha_i}(\rho)
\end{align}
\item Second term: $-V_i(\rho, \Omega^i_5)\sum_{K_j} \chi_{K_j}^{j,J \alpha_j}(\rho) \mathcal{Y}^{\ell_{\eta_j} \ell_{\lambda_j}}_{K_j}(\Omega^j_5)$ \\

The projection proceeds as with the previous term.

\begin{align}
&-\int [\mathcal{Y}^{\ell_{\eta_i} \ell_{\lambda_i}}_{K_i}(\Omega^i_5)]^* V_i(\rho, \Omega^i_5) \sum_{K_j} \chi_{K_j}^{j,J \alpha_j}(\rho) \mathcal{Y}^{\ell_{\eta_j} \ell_{\lambda_j}}_{K_j}(\Omega^j_5) d\Omega^i_5 = \notag \\
&-\sum_{K_j} \chi_{K_j}^{j,J \alpha_j}(\rho) \int [\mathcal{Y}^{\ell_{\eta_i} \ell_{\lambda_i}}_{K_i}(\Omega^i_5)]^* V_i(\rho, \Omega^i_5) \mathcal{Y}^{\ell_{\eta_j} \ell_{\lambda_j}}_{K_j}(\Omega^j_5) d\Omega^i_5
\end{align} 

The off-diagonal coupling potential between channels $i$ and $j$ is defined by
\begin{align}
  V^{ij}_{K_iK_j}(\rho) = \int [\mathcal{Y}^{\ell_{\eta_i} \ell_{\lambda_i}}_{K_i}(\Omega^i_5)]^* V_i(\rho, \Omega^i_5) \mathcal{Y}^{\ell_{\eta_j} \ell_{\lambda_j}}_{K_j}(\Omega^j_5) d\Omega^i_5  
\end{align}

Therefore the projected term is given by

\begin{align}
-\sum_{K_j} V^{ij}_{K_iK_j}(\rho) \chi_{K_j}^{j,J \alpha_j}(\rho)
\end{align}

\item Third term: $-V_i(\rho, \Omega^i_5) \sum_{K_k} \chi_{K_k}^{k,J \alpha_k}(\rho) \mathcal{Y}^{\ell_{\eta_k} \ell_{\lambda_k}}_{K_k}(\Omega^k_5)$

The projection is

\begin{align}
-\int [\mathcal{Y}^{\ell_{\eta_i} \ell_{\lambda_i}}_{K_i}(\Omega^i_5)]^* V_i(\rho, \Omega^i_5) \sum_{K_k} \chi_{K_k}^{k,J \alpha_k}(\rho) \mathcal{Y}^{\ell_{\eta_k} \ell_{\lambda_k}}_{K_k}(\Omega^k_5) d\Omega^i_5 = \notag \\
-\sum_{K_k} \chi_{K_k}^{k,J \alpha_k}(\rho) \int [\mathcal{Y}^{\ell_{\eta_i} \ell_{\lambda_i}}_{K_i}(\Omega^i_5)]^* V_i(\rho, \Omega^i_5) \mathcal{Y}^{\ell_{\eta_k} \ell_{\lambda_k}}_{K_k}(\Omega^k_5) d\Omega^i_5
\end{align}

Define the off-diagonal coupling potential between channels $i$ and $k$

\begin{align}
V^{ik}_{K_iK_k}(\rho) = \int [\mathcal{Y}^{\ell_{\eta_i} \ell_{\lambda_i}}_{K_i}(\Omega^i_5)]^* V_i(\rho, \Omega^i_5) \mathcal{Y}^{\ell_{\eta_k} \ell_{\lambda_k}}_{K_k}(\Omega^k_5) d\Omega^i_5
\end{align}

The projected potential is 
\begin{align}
 -\sum_{K_k} V^{ik}_{K_iK_k}(\rho) \chi_{K_k}^{k,J \alpha_k}(\rho)   
\end{align}
\end{enumerate}

Combining all three projected potential terms, the right side of the coupled hyperradial equation is obtained.

\begin{align}
-\sum_{K'_i} V^{ii}_{K_iK'_i}(\rho) \chi_{K'_i}^{i,J \alpha_i}(\rho) - \sum_{K_j} V^{ij}_{K_iK_j}(\rho) \chi_{K_j}^{j,J \alpha_j}(\rho) - \sum_{K_k} V^{ik}_{K_iK_k}(\rho) \chi_{K_k}^{k,J \alpha_k}(\rho)
\end{align}

This expression represents the sum of coupling terms from all three Faddeev partitions. To write this more compactly, a unified notation is introduced where the channel index $n$ runs over all three spectator configurations $n \in \{i,j,k\}$, and $K_n$ denotes the hyperangular momentum quantum number associated with partition $n$. The three separate sums can then be combined as

\begin{align} \label{eq:potentials}
- \sum_{n, K_n} V^{in}_{K_iK_n}(\rho) \chi_{K_n}^{n,J \alpha_n}(\rho)
\end{align}

For each value of $n$, the sum over $K_n$ runs over all hyperangular momentum values included in the basis for that partition. Combining the left-hand side of the projected Faddeev equations (Eq. \ref{eq:faddeev_coupled4}) and the projected right-hand side, one arrives at the following equation:

\begin{align}\label{eq:coupled_hyperradial_final}
 \left[-\frac{\hbar^2}{2m}\frac{d^2}{d\rho^2}  + L_{K_i} -E
   \right]    \chi_{K_i}^{i,J \alpha_i}(\rho) + \sum_{n,K_n} V^{in}_{K_iK_n}(\rho) \chi_{K_n}^{n,J \alpha_n}(\rho) = 0
\end{align}

\subsection{Raynal--Revai transformations}

In Eq. \ref{eq:coupled_hyperradial_final}, $V^{in}_{K_iK_n}$ is the coupling potential between different channels ($i, K_i$) and ($n, K_n$), where $n \in {i,j,k}$. These coupling matrix elements are given by $V^{in}_{K_iK_n}(\rho)= \langle \mathcal{Y}^{\ell_{\eta_i} \ell_{\lambda_i}}_{K_i}(\Omega^i_5) |V_i| \mathcal{Y}^{\ell_{\eta_n} \ell_{\lambda_n}}_{K_n}(\Omega^n_5) \rangle$. For $n=i,j,k$, the coupling potentials are as follows:

\begin{align}
V^{ii}_{K_iK'_i}(\rho)= \langle \mathcal{Y}^{\ell_{\eta_i} \ell_{\lambda_i}}_{K_i}(\Omega^i_5) |V_i| \mathcal{Y}^{\ell_{\eta_i} \ell_{\lambda_i}}_{K'_i}(\Omega^i_5) \rangle \\
V^{ij}_{K_iK_j}(\rho)= \langle \mathcal{Y}^{\ell_{\eta_i} \ell_{\lambda_i}}_{K_i}(\Omega^i_5) |V_i| \mathcal{Y}^{\ell_{\eta_j} \ell_{\lambda_j}}_{K_j}(\Omega^j_5) \rangle \\
V^{ik}_{K_iK_k}(\rho)= \langle \mathcal{Y}^{\ell_{\eta_i} \ell_{\lambda_i}}_{K_i}(\Omega^i_5) |V_i| \mathcal{Y}^{\ell_{\eta_k} \ell_{\lambda_k}}_{K_k}(\Omega^k_5) \rangle
\end{align}

For the case where $n=j$ and $n=k$, the hyperspherical harmonics $\mathcal{Y}^{\ell_{\eta_j} \ell_{\lambda_j}}_{K_j}(\Omega^j_5)$ and $\mathcal{Y}^{\ell_{\eta_k} \ell_{\lambda_k}}_{K_k}(\Omega^k_5)$ must be expressed in terms of  $\mathcal{Y}^{\ell_{\eta_i} \ell_{\lambda_i}}_{K_i}(\Omega^i_5)$, the hyperspherical harmonics corresponding to Jacobi partition $i$. This is done through the following transformations:

\begin{align}
\mathcal{Y}^{\ell_{\eta_j} \ell_{\lambda_j}}_{K_j}(\Omega^j_5) = \sum_{\ell_{\eta_i}, \ell_{\lambda_i}, K'_i} RR\langle \ell_{\eta_i} \ell_{\lambda_i} K'_i | \ell_{\eta_j} \ell_{\lambda_j} K_j \rangle_{\text{L}} \, \mathcal{Y}^{\ell_{\eta_i} \ell_{\lambda_i}}_{K'_i}(\Omega^i_5), 
\end{align}

\begin{align}
\mathcal{Y}^{\ell_{\eta_k} \ell_{\lambda_k}}_{K_k}(\Omega^k_5) = \sum_{\ell_{\eta_i}, \ell_{\lambda_i}, K'_i} RR\langle \ell_{\eta_i} \ell_{\lambda_i} K'_i | \ell_{\eta_k} \ell_{\lambda_k} K_k \rangle_{\text{L}} \, \mathcal{Y}^{\ell_{\eta_i} \ell_{\lambda_i}}_{K'_i}(\Omega^i_5), 
\end{align}

where $RR\langle \ell_{\eta_i} \ell_{\lambda_i} K'_i | \ell_{\eta_j} \ell_{\lambda_j} K_j \rangle_{\text{L}}$ and $RR\langle \ell_{\eta_i} \ell_{\lambda_i} K'_i | \ell_{\eta_k} \ell_{\lambda_k} K_k \rangle_{\text{L}}$ are Raynal-Revai coefficients \cite{ray1970, ers2016, kha1999, you1987,pup2009, pup1998, ray1973}. The Raynal–Revai coefficients depend on the total orbital angular momentum $L$, which is conserved under kinematic rotations between Jacobi partitions. Through this transformation, all the potential coupling matrix elements can be expressed in terms of hyperspherical harmonics for one Jacobi set. For the case $n=i$, the Raynal–Revai transformation reduces to the identity operator, i.e. a Kronecker delta in the hyperspherical harmonic basis. 

As outlined in the introduction, the quantum-mechanical three-body problem emerges across diverse fields, including nuclear physics, atomic and molecular physics, and particle physics. The formalism and equations derived here thus have wide-ranging applications in these domains. For computational implementation, a dedicated computer code for solving this system is detailed in \cite{tho2004}. For an in-depth exploration of numerical techniques for hyperspherical harmonics and the associated coupled hyperradial equations, refer to the review in \cite{kri1998}.

\subsection{Bound-state solutions}

In the hyperspherical harmonics method, the hyperradial wavefunctions $\chi_{K_i}^{i,J\alpha_i}(\rho)$ satisfy the system of coupled second-order differential equations in Eq. \ref{eq:coupled_hyperradial_final}, where the centrifugal term is  

\begin{align}
L_{K_i}(\rho) = \frac{\hbar^2}{2m} \frac{K_i(K_i+4) + 15/4}{\rho^2}.
\end{align}

To determine the behaviour of the solutions near the origin ($\rho \to 0$) and at infinity ($\rho \to \infty$), we analyze the dominant terms in these limits.

\subsubsection{Boundary condition at the origin ($\rho \to 0$)}

For small $\rho$, the centrifugal term $L_{K_i}(\rho) \propto 1/\rho^2$ diverges, while the potential terms $V_{K_i K_n}^{in}(\rho)$ typically remain finite or diverge more slowly, and the energy $E$ is negligible compared to $1/\rho^2$. Therefore, near $\rho = 0$, after ignoring negigible terms, Eq. \ref{eq:coupled_hyperradial_final} reduces to

\begin{align}
\left[ -\frac{\hbar^2}{2m} \frac{d^2}{d\rho^2} + \frac{\hbar^2}{2m} \frac{K_i(K_i+4) + 15/4}{\rho^2} \right] \chi_{K_i}^{i,J\alpha_i}(\rho) \approx 0.
\end{align}

Multiplying by $-2m/\hbar^2$ gives

\begin{align}
\label{eq:reduced_chi}
\frac{d^2 \chi}{d\rho^2} - \frac{\gamma}{\rho^2} \chi = 0,
\qquad
\gamma = K_i(K_i+4) + \frac{15}{4}.
\end{align}

where $\chi_{K_i}^{i,J\alpha_i}(\rho)$ is written simply as $\chi$. Eq. \ref{eq:reduced_chi} is a homogeneous second-order equation with variable coefficients, commonly called an equidimensional (or Cauchy--Euler) equation. It has a regular singular point at $\rho=0$ because of the coefficient of $\chi$ diverges as $1/\rho^2$ term. Such equations admit solutions given by a power-law $\chi(\rho) = \rho^q$. Substituting this solution into Eq. \ref{eq:reduced_chi}:

\begin{align}
q(q-1) \rho^{q-2} - \gamma \rho^{q-2} = 0.
\end{align}

Since $\rho^{q-2} \neq 0$ for $\rho > 0$, one obtains the indicial equation

\begin{align}
q(q-1) - \gamma = 0.
\end{align}

Now,

\begin{align}
\gamma = K_i(K_i+4) + \frac{15}{4}
      = K_i^2 + 4K_i + \frac{15}{4}.
\end{align}

The indicial equation becomes

\begin{align}
q^2 - q - \left( K_i^2 + 4K_i + \frac{15}{4} \right) = 0.
\end{align}

Solving this equation for $q$, one finds that the two roots are

\begin{align}
q_1= K_i + \frac{5}{2},
\qquad
q_2 = -\left( K_i + \frac{3}{2} \right).
\end{align}

The general solution near $\rho = 0$ is therefore

\begin{align}
\chi(\rho) \approx A \, \rho^{K_i + 5/2} + B \, \rho^{-(K_i + 3/2)}.
\end{align}

The second term diverges as $\rho \to 0$ for $K_i \ge 0$. To keep the wavefunction regular (finite at the origin), we must set $B = 0$. Hence, the regular solution behaves as

\begin{align}
\chi_{K_i}^{i,J\alpha_i}(\rho) \sim \rho^{K_i + 5/2} \quad \text{as} \ \rho \to 0.
\end{align}

Often, the factor $1/\rho^{5/2}$ is extracted in the expansion in Eq. \ref{eq:expansion1} to make the hyperradial equation Schr\"{o}dinger-like. If we define  

\begin{align}
\tilde{\chi}_{K_i}(\rho) = \frac{\chi_{K_i}(\rho)}{\rho^{5/2}},
\end{align}

then near the origin

\begin{align}
\tilde{\chi}_{K_i}(\rho) \sim \rho^{K_i}.
\end{align}

\subsubsection{Boundary condition at infinity ($\rho \to \infty$)}

For large $\rho$, the centrifugal term $L_{K_i}(\rho) \propto 1/\rho^2$ and the potential terms $V_{K_i K_n}^{in}(\rho)$ vanish (for short-range interactions). Equation \ref{eq:coupled_hyperradial_final} then reduces to

\begin{align}
\label{eq:reduced_chi2}
\left[ -\frac{\hbar^2}{2m} \frac{d^2}{d\rho^2} - E \right] \chi_{K_i}^{i,J\alpha_i}(\rho) = 0.
\end{align}

For bound states ($E < 0$), let

\begin{align}
\kappa = \sqrt{-\frac{2mE}{\hbar^2}} > 0.
\end{align}

Equation \ref{eq:reduced_chi2} becomes

\begin{align}
\frac{d^2 \chi}{d\rho^2} - \kappa^2 \chi = 0,
\end{align}

This is a second-order linear homogeneous ordinary differential equation with constant coefficients. From its characteristic equation, one finds its general solution to be

\begin{align}
\chi(\rho) = B_1 e^{-\kappa \rho} + B_2 e^{+\kappa \rho}.
\end{align}

The term $e^{+\kappa \rho}$ grows exponentially as $\rho \to \infty$ and is unphysical for a bound state. Thus, we set $B_2 = 0$, yielding

\begin{align}
\chi_{K_i}^{i,J\alpha_i}(\rho) \sim e^{-\kappa \rho} \quad \text{as} \ \rho \to \infty.
\end{align}

This exponential decay is characteristic of problems with potentials that decay as $\rho$ goes to infinity. For scattering states ($E > 0$), the solution is oscillatory, but that case is not treated here.

\section{Conclusion}

This paper presented a systematic treatment of the quantum-mechanical three-body problem. Its particular emphasis was on coordinate transformations and their role in simplifying the mathematical structure of the Schr\"odinger and Faddeev formalisms. Starting from the three-body Schr\"odinger equation in single-particle coordinates, mass-scaled Jacobi coordinates were introduced to separate internal and centre-of-mass dynamics. The transformation properties of gradients, Laplacians, and volume elements were derived explicitly using multivariable calculus, and the Jacobian determinants associated with coordinate transformations were computed to ensure mathematical consistency and preservation of normalization.

The kinetic energy operator was transformed step by step into Jacobi coordinates, demonstrating the emergence of a diagonal structure and the complete separation between internal and centre-of-mass motion. This separation provided a natural framework for formulating both the Schr\"odinger equation and the Faddeev equations in a form suited to few-body calculations. The Faddeev formalism was then presented
in configuration space, highlighting its advantages in avoiding overcounting of interactions, simplifying boundary conditions, and isolating singular structures in realistic interaction potentials. Symmetry considerations for systems with indistinguishable particles were also discussed, leading to reduced coupled systems or single equations in highly symmetric cases.

The formulation was subsequently extended to Delves' hyperspherical coordinates, where the introduction of the hyperradius, one hyperangle and two sets of spherical polar angles allows a compact representation of the internal dynamics in a six-dimensional space. The associated Jacobians and volume elements were derived explicitly. This provided the necessary groundwork for hyperspherical harmonics expansions and the derivation of the widely used coupled hyperradial equations.

Overall, this study provides a coherent mathematical bridge from single-particle descriptions to modern few-body methods based on Jacobi and hyperspherical coordinates. The explicit derivations presented here are intended to serve both as a reference for researchers and as a pedagogical resource for graduate students entering the field of few-body physics. 

\section*{Acknowledgement}

The author is grateful to Professor Lorenzo Fortunato (Universit\`{a} di Padova) for an enlightening discussion on Jacobi coordinates.
\bibliographystyle{unsrtnat}
\bibliography{threebody_meoto}
\end{document}